\documentclass[pra,twocolumn,preprintnumbers,amsmath,amssymb,superscriptaddress]{revtex4}

\usepackage{color}
\usepackage{graphicx}
\usepackage{dcolumn}
\usepackage{bm}
\usepackage{hyperref}
\usepackage{enumitem}
\usepackage{balance}
\usepackage{comment}
\usepackage{cleveref}

\usepackage{amsthm}

\begin{document}
\title{Multipartite Spin Coherent States and Spinor States}

\author{Tim Byrnes}
 \email{tim.byrnes@nyu.edu}
\affiliation{New York University Shanghai, NYU-ECNU Institute of Physics at NYU Shanghai, Shanghai Frontiers Science Center of Artificial Intelligence and Deep Learning, 567 West Yangsi Road, Shanghai, 200126, China.}
\affiliation{State Key Laboratory of Precision Spectroscopy, School of Physical and Material Sciences, East China Normal University, Shanghai 200062, China}
\affiliation{Center for Quantum and Topological Systems (CQTS), NYUAD Research Institute, New York University Abu Dhabi, UAE.}
\affiliation{Department of Physics, New York University, New York, NY 10003, USA}

\begin{abstract}
Multipartite generalizations of spin coherent states are introduced and analyzed.  These are the spin analogues of multimode optical coherent states as used in continuous variable quantum information, but generalized to possess full spin symmetry.  
Two possible generalizations are given, one which is a simple tensor product of a given multipartite quantum state.  The second generalization uses the bosonic formulation in the Jordan-Schwinger map, which we call spinor states.  In the unipartite case, spinor states are equivalent to spin coherent states, however in the multipartite case, they are no longer equivalent.  Some fundamental properties of these states are discussed, such as their observables and covariances with respect to symmetric operators, form preserving transformations, and entanglement. We discuss the utility of such multipartite spin coherent and spinor states as a way of storing quantum information.  
\end{abstract}

\date{\today}

\maketitle

\section{Introduction}

The discovery of coherent states by Schrodinger \cite{schrodinger1926stetige} and its application to the quantum theory of light by 
Glauber \cite{glauber1963quantum} is one of the fundamental results in quantum optics.  Coherent states are minimal uncertainty states that  follow similar dynamics to the classical harmonic oscillator.  As the amplitude of the coherent state is increased, the quantum noise relative to the amplitude decreases \cite{scully1999quantum}, giving an example of the correspondence principle.  Their practical use stems from the fact that they are the idealized quantum state that emerges from a laser \cite{scully1999quantum}. They are the starting point for examining more complex states such as squeezed states, where the quantum fluctuations of one variable can be exchanged to another \cite{walls1983squeezed}.  By extending the system to the multimode context, entangled states of light may be produced which possess Einstein-Podolsky-Rosen (EPR) correlations \cite{heidmann1987observation,andersen201630,cavalcanti2009experimental}. Using such  multimode systems has been the basis for various applications such as quantum cryptography \cite{ralph1999continuous,grosshans2002continuous,wang2007quantum,xiang2017multipartite} and continuous variable quantum information \cite{braunstein2005quantum,adesso2007entanglement}.

While coherent states are naturally occurring states in optical systems, in other systems involving an ensemble of identical finite dimensional quantum particles, spin coherent states are their natural counterpart \cite{radcliffe1971some,arecchi1972atomic,combescure2012spin}. 
One major difference between the two types coherent states is that optical coherent states do not have a definite particle number.  Meanwhile, spin coherent states in the ideal case have a fixed particle number.  Experimentally, spin coherent states are often realized in atomic ensembles, where the internal atomic states form the spin ensemble \cite{hammerer2010quantum,gross2012spin}. 
Creating a spin coherent state is typically the first step in generating spin squeezing \cite{ma2011quantum}.  
In the case of an ensemble of two-level atoms, if they are polarized in a particular spin direction, the remaining two spins may be approximated by canonical position and momentum operators via the Holstein-Primakoff transformation \cite{holstein1940field,byrnes2021quantum}. In this way spin coherent states mimic optical coherent states, however, it should be kept in mind that this is a limiting case and spin ensembles offer richer physics due to their spin symmetries.  One and two-axis spin squeezing  was proposed \cite{wineland1992spin,kitagawa1993squeezed} and realized \cite{hald1999spin} with atomic ensembles. The primary application of such squeezed atomic ensembles to date has been for quantum metrology, where squeezed variables allow for measurements beyond the standard quantum limit \cite{ma2011quantum,gross2012spin}. 

Most of the theory and experiments relating to spin coherent states have focused upon single atomic ensembles.   Experiments involving more than one atomic ensemble have been performed within the Holstein-Primakoff approximation, where two-mode squeezed state generation was realized \cite{julsgaard2001experimental,krauter2013deterministic}.  The regime beyond the Holstein-Primakoff approximated regime has been examined primarily theoretically to date, extending notions of spin squeezing to two ensembles, such as with one and two-axis two-spin squeezed states \cite{byrnes2013fractality,kitzinger2020two}.  Using multiple spin ensembles has been considered in the context of several quantum information applications, such as remote state preparation \cite{chaudhary2021remote}, quantum teleportation \cite{pyrkov2014quantum,pyrkov2014full}, adiabatic quantum computing \cite{mohseni2021error}, and other quantum algorithms \cite{byrnes2012macroscopic,byrnes2015macroscopic,semenenko2016implementing}.  Entanglement between two split BECs was recently experimentally realized in Ref. \cite{colciaghi2023einstein}.  

In this paper, we 
examine the multipartite generalization of spin coherent states.  There are two formulations of spin coherent states which we show give different generalizations.  For example, we may write a unipartite (i.e. single ensemble) spin coherent state consisting of qubits as
\begin{align}
| \psi \rangle^{\otimes N } = \left( \alpha | 0 \rangle + \beta | 1 \rangle \right)^{\otimes N} ,
\label{qubitscsintro}
\end{align}
where $ \alpha, \beta $ are normalized amplitudes and $ N $ is the number of qubits.  This may be equivalently be written
\begin{align}
| \psi \rangle \rangle = \frac{1}{\sqrt{N!}} \left( \alpha a^\dagger + \beta b^\dagger \right)^{\otimes N} | \text{vac} \rangle 
\label{singlespinor}
\end{align}
which we will call a {\it spinor state} in this paper.  Here $ a, b $ are bosonic annihilation operators and $ | \text{vac} \rangle $ is the vacuum state.  For the unipartite case, spin coherent states and spinor states have a exact mathematical equivalence, hence thus far there has not been a need to distinguish them.  However, in the multipartite case, we show that spin coherent states and spinor states are no longer equivalent.  Spin coherent states generalize to the multipartite case in a straightforward way, where we write some of their basic properties in Sec. \ref{sec:multipartite}.  The basic difference 
between the two classes of states is the type of symmetry that they follow under particle interchange.  After showing some general properties of spinor states in Sec. \ref{sec:multipartitespinor}, we show an elementary example of a bipartite spinor state in Sec. \ref{sec:bipartite}. We discuss the potential applications of such states in Sec. \ref{sec:error} and methods to prepare them in Sec. \ref{sec:preparation}.


\section{Unipartite spin coherent and spinor states}
\label{sec:unipartite}



\subsection{Definitions}

We first review unipartite spin coherent states and some of their properties \cite{radcliffe1971some,arecchi1972atomic,combescure2012spin,byrnes2021quantum}.  Consider an $ L $-level quantum system, where the orthogonal states are labelled as $ |l \rangle $ with $ l \in [0,L-1] $. The single quantum system can be in an arbitrary superposition state, which we write as
\begin{align}
    |\psi\rangle = \sum_{l=0}^{L-1} \psi_l | l \rangle  .
    \label{singlestate}
\end{align}
Now consider $ N $ duplicates of this quantum system.  The total state of this system, the unipartite spin coherent state, can be written as 
\begin{align}
 |\psi\rangle^{\otimes N} = \left( \sum_{l=0}^{L-1} \psi_l | l \rangle  \right)^{\otimes N} . 
\label{scsdef}
\end{align}
The $ N $ duplicates of the quantum system have the same dimension and are prepared in the same quantum state. Physically, the spin coherent states can be realized in any system where many duplicate quantum systems are available.  A typical realization of (\ref{scsdef}) is an atomic ensemble, where the state of each atom is (\ref{singlestate}).  Considering the number of duplicates to be typically a large number $ N \gg 1 $, in this paper will call the unduplicated state (\ref{singlestate}) to be the {\it microscopic} state, while (\ref{scsdef}) is the {\it macroscopic} state.  We will defer further details of the physical implementation to Sec. \ref{sec:preparation}, and focus on the fundamental properties of such spin coherent states.  

The spin coherent state (\ref{scsdef}) is a completely symmetric state under particle interchange.  To see this, write (\ref{scsdef}) as
\begin{align}
 |\psi\rangle^{\otimes N} = \sum_{l_1,l_2 \dots, \l_N=0}^{L-1} \psi_{l_1}  \psi_{l_2} \dots \psi_{l_N} |l_1 l_2 \dots l_N \rangle .
 \label{expandedscs}
\end{align}
Exchanging the labels $ l_n \leftrightarrow l_{n'}$ for $ n, n' \in [1, N] $ leaves the wavefunction unchanged.  For this reason, it is possible to write a mathematically equivalent formulation of spin coherent state using bosonic operators \cite{byrnes2021quantum}.  The equivalent state to (\ref{scsdef}) in the bosonic formulation is
\begin{align}
| \psi \rangle \rangle =\frac{1}{\sqrt{N!}}  \left( \sum_{l=0}^{L-1} \psi_l a_l^\dagger \right)^N | \text{vac} \rangle  , 
\label{scsdefboson}
\end{align}
where $ [ a_l, a_{l'}^\dagger ] = \delta_{l l'} $ and the vacuum state satisfies $ a_l  | \text{vac} \rangle  = 0 $.  The state (\ref{scsdefboson}) is an elementary example of a spinor state. To distinguish the spinor state to the spin coherent state, we have used the ``double-ket'' notation \cite{byrnes2012macroscopic} which implies a $ N $-fold bosonic duplication.  There is no mathematical difference between a single or double-ket, it is purely for notational convenience.  However, the double-ket is suggestive of the fact that the state is a macroscopic quantum state for large $ N $.  

The spin coherent state (\ref{scsdef}) and spinor state (\ref{scsdefboson}) have a similar appearance but in fact are not necessarily equivalent, as we will see later.  For now, we point out that the Hilbert space that the states exists within are rather different. The bosonic formulation results in a considerable reduction in the dimension of the Hilbert space. To see this, expand (\ref{scsdefboson}) as (see Appendix \ref{app:equiv})
\begin{align}
|\psi \rangle \rangle = & \sum_{k_0 = 0}^N \dots \sum_{k_{L-1} = 0}^N \sqrt{\binom{N}{k_0, \dots, k_{L-1}} } \psi_0^{k_0} \dots \psi_{L-1}^{k_{L-1}} \nonumber \\
&  \times | k_0, \dots, k_{L-1} \rangle  .
\label{unispinorexpan}
\end{align}
where the normalized Fock states are
\begin{align}
|k_0, \dots, k_{L-1} \rangle = \frac{ (a_0^\dagger)^{k_0} \dots (a_{L-1}^\dagger)^{k_{L-1}}}{\sqrt{k_0! \dots k_{L-1}! }} | \text{vac} \rangle . 
\label{multifockone}
\end{align}
with $ \sum_{l=0}^{L-1} k_l = N$. Comparing the wavefunctions (\ref{unispinorexpan}) and (\ref{expandedscs}) it is possible to find an equivalence between the two states as discussed in Appendix \ref{app:equiv}. The primary difference between the two is that spin coherent states involve underlying atoms that are distinguishable in principle, whereas spinor states involve indistinguishable particles.  
Concretely, for the spin coherent state (\ref{scsdef}), the dimension of the Hilbert space is $ L^N $.  In comparison, the dimension of the space of (\ref{unispinorexpan}) is
\begin{align}
    D(N,L) = \binom{N+L-1}{L-1} . 
    \label{dimbosonic}
\end{align}
Due to the elimination of states which are not completely symmetric under particle interchange, the dimension of the bosonic formulation is considerably smaller.  For example, in an ensemble of two dimensional atoms ($ L = 2$), the Hilbert space dimension of (\ref{qubitscsintro}) is $ 2^N$, while the bosonic version (\ref{singlespinor}) has a dimension $ N + 1$.  

In addition to a mathematical equivalence to spin coherent states, the spinor states  (\ref{scsdefboson}) can be physically created from systems involving indistinguishable bosonic particles, such as in a spinor Bose-Einstein condensate (BEC).  Typically, in this case, $ a_l  $ denote the annihilation operators for the atoms in different internal states.  Since all other degrees of freedom (e.g. spatial degrees of freedom) are identical, the atoms in the BEC all occupy the same physical state given by (\ref{singlestate}).  Such states are also referred to as spinor BEC \footnote{The name {\it spinor} arises from the fact that they form a representation of the SU($L$) group, where the $ N $ labels the particular representation. Here we will not necessarily follow this definition for our spinor states, and use it more loosely as any state that follows the form (\ref{multiscsbosonic}).}.

\subsection{Example: $ L= 2 $ level system} 

To illustrate the above, consider the $ L = 2 $ level case, where the spin coherent state is written
\begin{align}
|\theta, \phi \rangle^{\otimes N} = \left( \cos \frac{\theta}{2} |0 \rangle  + e^{i \phi} \sin \frac{\theta}{2} |1 \rangle \right)^{\otimes N}  ,
\label{qubitscs}
\end{align}
where the equivalent spinor version is 
\begin{align}
| \theta, \phi \rangle \rangle = \frac{1}{\sqrt{N!}} \left( \cos \frac{\theta}{2} a^\dagger + e^{i \phi} \sin \frac{\theta}{2} b^\dagger \right)^N |\text{vac} \rangle ,
\label{qubitspinor}
\end{align}
where $ \theta \in [0,\pi] $ and $ \phi \in [0, 2 \pi ] $. The bosonic operators satisfy commutation relations $ [a, a^\dagger ] = [ b, b^\dagger ] = 1$ and $ [a,b]=0 $.  In this case, the degrees of freedom of the spin coherent state are identical to that of a qubit, since the spin coherent state and spinor state transforms under the spin-$N/2$ representation of SU(2). 

The spinor nature of the state may be seen by examining the total spin operators
\begin{align}
    S^x & = \sum_{n=1}^N \sigma^x_n \nonumber \\
    S^y & = \sum_{n=1}^N \sigma^y_n \nonumber \\
    S^z & = \sum_{n=1}^N \sigma^z_n  ,  
    \label{totalspin}
\end{align}
where $ \sigma^j_n $ is a Pauli spin operator for the $ n$th duplicate qubit.  These obey the commutation relations $ [S^j, S^k ] = 2 i \epsilon_{jkl} S^l $, where $ \epsilon_{jkl} $ is the Levi-Civita antisymmetric tensor.  The total spin operator can also be written in the bosonic formulation
\begin{align}
\label{schwingerboson}
S^x & = a^\dagger b + b^\dagger a, \nonumber\\
S^y & = -ia^\dagger b + ib^\dagger a, \nonumber\\
S^z & = a^\dagger a - b^\dagger b.
\end{align}
We use the same notation for the spin coherent state and spinor state since it will be self-evident which version should be used in each case, and they have similar properties in most cases.    

The expectation values of the total spin operators in either formulation are
\begin{align}
\langle S^x \rangle & = N \sin \theta \cos \phi  \nonumber \\
\langle S^y \rangle & = N \sin \theta \sin \phi  \nonumber \\
 \langle S^z \rangle & = N \cos \theta 
\end{align}
which are the same as for a qubit, but multiplied by $ N $.  We may thus understand the spin coherent state to be a polarized state where all the qubits point in the same spin direction.  This shows the homomorphism between SU(2) to SO(3), which characterizes spinors \cite{steane2013introduction}.  

The spin coherent state (\ref{qubitscs}) and spinor state (\ref{qubitspinor}) are eigenstates of the Hamiltonian
\begin{align}
    H = - \bm{n} \cdot \bm{S}
\end{align}
where $ \bm{n} = (n_x, n_y, n_z ) $ and
\begin{align}
n_x & = \sin \theta \cos \phi \nonumber \\
n_y & = \sin \theta \sin \phi \nonumber \\
n_z & = \cos \theta .
\end{align}
They are the ground states of their respective Hamiltonians with eigenvalue $ - N $.  For example, for the spinor state,
\begin{align}
    H | \theta, \phi \rangle \rangle = - N | \theta, \phi \rangle \rangle ,
\end{align}
and similarly for the spin coherent state.

\section{Multipartite spin coherent states}
\label{sec:multipartite}

\subsection{Definition}

Constructing the spin coherent state as in (\ref{scsdef}) or (\ref{scsdefboson}) follows a simple procedure: one starts with a single quantum system and duplicates it $ N $ times.  While the spin coherent state is already a many-body state involving $ N $ quantum atoms, in terms of the quantum information stored on the state, it is equivalent to a single quantum system, since the $ N $ quantum systems are duplicates.  For example, the spin coherent state (\ref{qubitscs}) is parameterized by two variables $ \theta, \phi $, which is the same as a pure state of a single qubit.  

We now generalize spin coherent states to the multipartite case.  We define the multipartite spin coherent state as
\begin{align}
| \Psi \rangle^{\otimes N} = \left(
\sum_{l_1, \dots, l_M} \Psi_{l_1 \dots l_M} |l_1 \rangle \otimes | l_2 \rangle \otimes \dots \otimes | l_M \rangle  \right)^{\otimes N} . \label{multiscs}
\end{align}
Here, there underlying multipartite state involves $ M $ subsystems with basis states $  |l_1 \rangle \otimes | l_2 \rangle \otimes \dots \otimes | l_M \rangle  $.  As before, the multipartite state is duplicated $ N $ times. 

\begin{figure}[t]
\includegraphics[width=\linewidth]{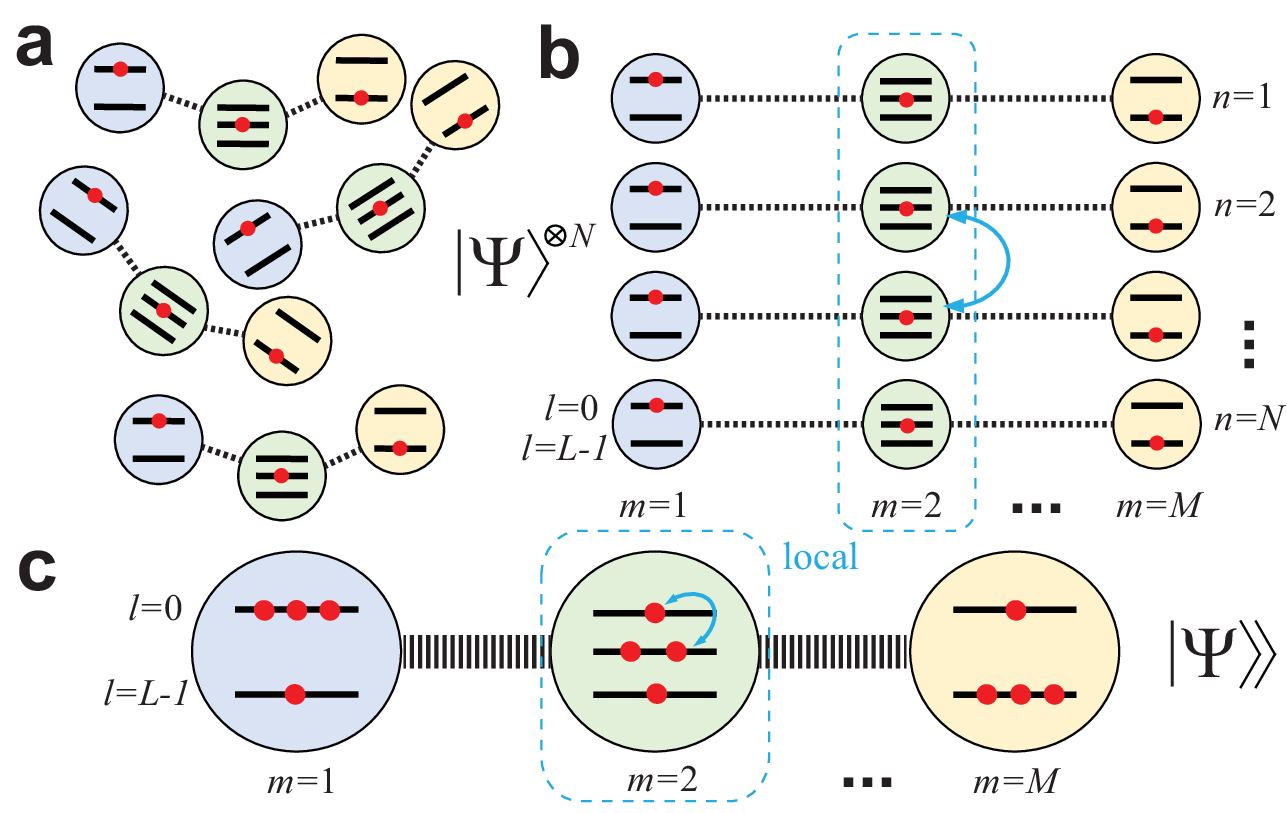}
\caption{Multipartite spin coherent states and spinor states as considered in this paper. Spin coherent states (\ref{multiscs}) are shown in (a) and (b), while spinor states (\ref{multiscsbosonic}) are shown in (c). 
Each of the $ N $ ``molecules'' are in the state $ | \Psi \rangle $.  (b) shows a rearranged version of (a) and is the identical quantum state.  The labelling conventions in the paper are shown.  Thick horizontal lines in the circles are distinct quantum states and the dot indicates occupancy of the state.  Dotted lines indicate entanglement between the subsystem.  The figure shows the case of $ M = 3$ subsystems, $ N =4$ duplicates, and subsystem dimension $L = 2,3 $ (qubit or qutrit).  The dashed box in (b)(c) show the notion of locality.  Arrows show examples of local particle interchange.  Spinor states are symmetric under local particle interchange (subfigure (c)), while spin coherent states (subfigure (b)) are not. 
\label{fig1}  }
\end{figure}

The physical configuration corresponding to (\ref{multiscs}) can be pictured in Fig. \ref{fig1}(a).  
Physically, this may be realized by preparing $ N $ duplicate quantum systems, each consisting of $ M $ subsystems.  In this paper, we call each of the duplicates a {\it molecule}, as suggested by the molecular gas picture shown in Fig. \ref{fig1}(a).  Each molecule is in a state  $ |\Psi\rangle $, and the full system consists of an ensemble of such molecules.  Within each molecule, each of the  $M $ subsystems do not necessarily need to be the same dimension $L $.  For example, in Fig. \ref{fig1}(a), the molecule consists of two qubits and a qutrit.  In most cases that we consider in this paper, the dimension $ L $ is the same for all subsystems. 

We now define the notion of locality for  the multipartite spin coherent state.  Consider rearranging in the configuration of the molecules in Fig. \ref{fig1}(a) into that of Fig. \ref{fig1}(b), where each of the subsystems of the multipartite wavefunction are grouped together.  The structure of the entanglement still remains the same, where each particle is entangled to its corresponding subsystem.  We call any operation that only deals with the $ m $th subsystem ($ m\in [1,M] $) a {\it local} operation. Such a definition of locality gives rise to non-trivial quantum correlations, due to the presence of entanglement between the subsystems.  


%

\subsection{Expectation values and covariances}
\label{sec:expcov}

Now let us evaluate the expectation values and variances of the multipartite spin coherent state. We will primarily consider observables that are symmetric with respect to particle interchange in a similar way to the spin operators (\ref{totalspin}) taking a form
\begin{align}
C = \sum_{n=1}^N c_{n} ,  
\label{symmetricop}
\end{align}
where $ c_{n} $ is an operator that acts on the $ n $th molecule of the spin coherent state, with $ n \in [1,N] $.  We consider such operators that are symmetric under particle interchange based on a physical motivation, that one typically cannot access a particular microscopic molecule in an ensemble.  For example, in an atomic/molecular gas ensemble, one may be able to access the global properties of the ensemble, but not a single atom/molecule of the ensemble.  

For such symmetric operators, there is a simple relationship between the microscopic and macroscopic expectation values.  For a unipartite expectation value we have
\begin{align}
\langle C \rangle & \equiv \langle \Psi |^{\otimes N } C | \Psi \rangle^{\otimes N } \nonumber \\
& = N \langle c \rangle 
\label{uniexp}
\end{align}
where 
\begin{align}
\langle c \rangle  = \langle \Psi | c_{n} | \Psi \rangle .  
\label{singlepartexp}
\end{align}
Here, we have dropped the $n$ label because any of the $N $ duplicate molecular states are the same. 

Expectation values of products of symmetric operators generally give more complex relations.  For example for two operators $ C $ and $ D=\sum_{n=1}^N d_n $ taking the same form as (\ref{symmetricop}), we have
\begin{align}
\langle CD \rangle = N \langle cd \rangle + N(N-1) \langle c \rangle \langle d \rangle ,  
\end{align}
where 
\begin{align}
\langle c d \rangle = \langle \Psi| c_n  d_n | \Psi \rangle . 
\end{align}
The contribution from the microscopic correlations are dominated by the single particle expectation values of order $ N^2$.  Covariances, on the other hand, have a simpler relation.  For example, we have
\begin{align}
\text{Cov} (C, D) & = \langle C D \rangle - \langle C \rangle \langle D \rangle \nonumber \\
& = N ( \langle c d \rangle - \langle c \rangle \langle d \rangle ) = N \text{Cov} (c, d)
\label{bipartiteexp}
\end{align}
The simple relation arises from a neat cancellation of terms of order $ N^2 $.  In this sense, covariances are the natural quantity to relate the microscopic and macroscopic quantities involving products of operators.  It follows that variances have the same relation
\begin{align}
\text{Var}(C) & = \langle C^2 \rangle - \langle C \rangle^2 \nonumber \\
& = N( \langle c^2 \rangle - \langle c \rangle^2 ) =
N \text{Var}(c) .
\label{variancec}
\end{align}

The above may be straightforwardly be generalized to higher order correlators.  Specifically, central moments relate macroscopic and microscopic quantities.  For example, the third order central moments are related as
\begin{align}
& \langle (C- \langle C \rangle)(D- \langle D \rangle)(E- \langle E \rangle) \rangle \nonumber \\
& = N \langle (c- \langle c \rangle)(d- \langle d \rangle)(e- \langle e \rangle) \rangle .
\end{align}
where $ E=\sum_{n=1}^N e_n $ is a symmetric operator.


\subsection{Form preserving transformations}
\label{sec:scspres}

The multipartite spin coherent state is a subclass of states that has the specific form (\ref{multiscs}). Under a general unitary transformation, a spin coherent state will not in general remain a spin coherent state.  It is therefore of interest to know what transformations leave the state in the class of states (\ref{multiscs}).  

For the multipartite spin coherent states this may be easily answered. For the states of the form (\ref{multiscs}), clearly any unitary transformation of the form
\begin{align}
U & = u^{\otimes N} = \prod_{n=1}^N e^{-i t H_n/\hbar } = e^{-\frac{it}{\hbar} \sum_n H_n} ,
\label{presevingunitary}
\end{align}
where $ u $ is a unitary operator acting on the $ n$th duplicate, $ H_n $ is the underlying Hamiltonian for this operation, and $ t $ is the evolution time. The above unitary simply transforms all the duplicates in the system in the same way, such that all the final states are all the same state
\begin{align}
U | \Psi \rangle^{\otimes N} = | \Psi' \rangle^{\otimes N} ,
\end{align}
where $ | \Psi' \rangle  = u | \Psi \rangle $ is the microscopic transformation for each molecule.  

A sufficient condition for the preservation of a spin coherent state is then that the Hamiltonian of the associated unitary transformation has form that is symmetric under interchange of the molecules
\begin{align}
H = \sum_n H_n ,
\label{scsham}
\end{align}
which is evident from (\ref{presevingunitary}).

\section{Multipartite spinor states}
\label{sec:multipartitespinor}

\subsection{Definition}

We now make another generalization of the spin coherent state, this time using the bosonic formulation (\ref{scsdefboson}).  The multipartite spinor state is defined as
\begin{align}
| \Psi \rangle \rangle = \frac{1}{\sqrt{{\cal N}_\Psi}}
\left(
\sum_{l_1, \dots, l_M} \Psi_{l_1 \dots l_M} a^\dagger_{1,l_1} \dots 
 a^\dagger_{M,l_M}  \right)^{N}  | \text{vac} \rangle  ,
 \label{multiscsbosonic}
\end{align}
where $ a_{m,l} $ labels a bosonic annihilation operator for a boson in the $ m $th subsystem, in the $l$th state.  The wavefunction $ \Psi_{l_1 \dots l_M} $ is the same wavefunction that appears in the multipartite spin coherent state (\ref{multiscs}). While  $ \Psi_{l_1 \dots l_M} $ is a normalized wavefunction, the spinor state will not necessarily be normalized, hence we include a normalization factor $ {\cal N }_\Psi $.  The normalization factor is not a universal constant as we saw in the unipartite case, and is dependent on the particular state $ \Psi $.  We remind the reader of this dependence with a subscript  $ {\cal N }_\Psi $.  

To understand what kind of a state the spinor state is, let us write the states on the $m$th subsystem.  The normalized Fock states are defined as
\begin{align}
|k_0, \dots, k_{L-1} \rangle_m = \frac{ (a_{m,0}^\dagger)^{k_0} \dots (a_{m,L-1}^\dagger)^{k_{L-1}}}{\sqrt{k_0! \dots k_{L-1}! }} | \text{vac} \rangle ,  
\label{multifock}
\end{align}
where $ \sum_{l=0}^{L-1} k_l = N $.  This type of state can be pictured in the way as shown in Fig. \ref{fig1}(c).  The $ N $ bosons on each subsystem can be distributed among the $ L $ levels, where each level can be occupied by more than one boson.  A difference that is immediately apparent here is that the identity of the unduplicated state is less easily seen in the spinor case compared to the spin coherent state.  Despite the relatively simple form of the wavefunction (\ref{multiscsbosonic}), its underlying state is ultimately a entangled state of $ M $ lots of $ D(N, L) $ dimensional qudits, as given in (\ref{dimbosonic}).  
 For example, for a $ L = 2$ subsystem, the dimension of the full system is $ (N + 1)^M$.

\subsection{Form preserving transformations}
\label{sec:formspinor}

In a similar way to multipartite spin coherent states, we examine unitary transformations that preserve the form of the spinor states.  Namely, we look for unitary transformations that realize $ | \Psi \rangle \rangle \rightarrow | \Psi ' \rangle \rangle $.

In contrast to the spin coherent state case where it is possible to write down a form preserving unitary for general unitary transformations $ | \Psi \rangle^{\otimes N } \rightarrow | \Psi' \rangle^{\otimes N } $, for spinor states, this is more difficult due to the collective nature of subsystems.  However, {\it local} linear unitary transformations are form preserving. Consider a transformation of the form
\begin{align}
V_m = \exp(-i \sum_{l l'} H_{l l'} a^\dagger_{m,l} a_{m,l'} )
\label{localrot}
\end{align}
where 
$ H_{l l'}$ is a Hermitian matrix. 
Applying this to the spinor state  we have
\begin{align}
& V_1 | \Psi \rangle \rangle = \frac{V_1}{\sqrt{{\cal N}_\Psi}} 
\left(
\sum_{l_1, \dots, l_M} \Psi_{l_1 \dots l_M} a^\dagger_{1,l_1} \dots 
 a^\dagger_{M,l_M}  \right)^{N} \nonumber \\
&  \times V_1^\dagger V_1 | \text{vac} \rangle \nonumber \\
& = \frac{1}{\sqrt{{\cal N}_\Psi}} 
\left(
\sum_{l_1, \dots, l_M} \Psi_{l_1 \dots l_M} V_1 a^\dagger_{1,l_1}V_1^\dagger \dots 
 a^\dagger_{M,l_M}  \right)^{N} | \text{vac} \rangle \nonumber \\
 & = \frac{1}{\sqrt{{\cal N}_\Psi}} 
\left(
\sum_{l_1, \dots, l_M} \sum_{l_1'} \Psi_{l_1 \dots l_M}  v_{l_1 l_1'} a^\dagger_{1,l_1'} \dots 
 a^\dagger_{M,l_M}  \right)^{N} | \text{vac} \rangle 
\end{align}
where we have chosen $ m =1 $ without loss of generality and used the fact that a linear unitary transformation restricted to the $m$th subsystem linearly transforms the bosonic operators according to 
\begin{align}
V_m a^\dagger_{m,l} V_m^\dagger = \sum_{l'} v_{l l'} a^\dagger_{m,l'} .  
\end{align}
Hence we conclude that applying any local operator of the form (\ref{localrot}) will preserve the spinor nature of the state
\begin{align}
V_m | \Psi \rangle \rangle = | \Psi' \rangle \rangle ,
\end{align}
where the transformation of the bosonic operators works in exactly the same way as a local transformation for the underlying state.  That is, for the underlying state
\begin{align}
|\Psi \rangle = \sum_{l_1, \dots, l_M} \Psi_{l_1 \dots l_M} |l_1 , \dots, l_M \rangle   
\end{align}
applying the local operation on the $ m$th subsystem 
\begin{align}
V_m = \exp(-i \sum_{l l'} H_{l l'} |l\rangle_m \langle l' |_m  )
\end{align}
gives the state 
\begin{align}
|\Psi ' \rangle = V_m | \Psi \rangle . 
\end{align}
Clearly, multiple applications of local transformations also preserve the spinor state form.

\subsection{Inequivalence to spin coherent states}

Some of the above aspects, such as the state-dependent normalization and the lack of a general form preserving transformation already suggests that the spinor state is in fact inequivalent to spin coherent states in the multipartite case.  Here we explicitly show the reason for this inequivalence. 

We show the inequivalence by contradiction, by explicitly trying to establish a mapping in a similar way to that done in Appendix \ref{app:equiv}.  Considering the $ M=2, L=2, N=2 $ case, write the spin coherent state
\begin{align}
& \left( \alpha| 00 \rangle + \beta | 01 \rangle + \gamma | 10 \rangle + \omega | 11 \rangle   \right)^{\otimes 2} \nonumber \\
& = \alpha^2 | 00 \rangle | 00 \rangle + \beta^2 | 01 \rangle | 01 \rangle + \gamma^2 | 10 \rangle | 10 \rangle + \omega^2 | 11 \rangle| 11 \rangle \nonumber \\
& + \alpha \beta ( | 00 \rangle | 01 \rangle + | 01 \rangle |00 \rangle )  + \alpha \gamma ( | 00 \rangle | 10 \rangle + | 10 \rangle |00 \rangle ) \nonumber \\
& + \beta \omega ( | 01 \rangle | 11 \rangle + | 11 \rangle | 01 \rangle ) + \gamma \omega ( | 10 \rangle | 11 \rangle + | 11 \rangle | 10 \rangle ) \nonumber \\
& + \alpha \omega ( | 00 \rangle | 11 \rangle + | 11 \rangle |00 \rangle ) + \beta \gamma ( | 01 \rangle | 10 \rangle + | 10 \rangle | 01 \rangle  ) ,
\label{scsn2ex}
\end{align}
where $ \alpha, \beta, \gamma, \omega $ are  normalized complex coefficients. 
We may see how the form of the spin coherent state makes each of the terms with the same coefficient to be symmetrized under a bipartite interchange.  In the general case, spin coherent states involve Fock states that are symmetric under a $ M$-particle interchange.  

Compare this to the unnormalized spinor state with the same coefficients 
\begin{align}
& ( \alpha a^\dagger_1 a^\dagger_2 + \beta a^\dagger_1 b^\dagger_2
+ \gamma b^\dagger_1 a^\dagger_2 + \omega b^\dagger_1 b^\dagger_2)^2 | \text{vac} \rangle \nonumber \\
= \Big[ & \alpha^2 (a^\dagger_1 a^\dagger_2)^2 + \beta^2 (a^\dagger_1 b^\dagger_2 )^2 + \gamma^2 (b^\dagger_1 a^\dagger_2)^2 +\omega^2 (b^\dagger_1 b^\dagger_2)^2 \nonumber \\
& + 2 \alpha \beta ( a^\dagger_1)^2 a^\dagger_2 b^\dagger_2
+ 2 \alpha \gamma a^\dagger_1 b^\dagger_1 (a^\dagger_2)^2 +  \beta  \omega  a^\dagger_1 b^\dagger_1 (b^\dagger_2)^2 \nonumber \\
& 
+ \gamma  \omega  (b^\dagger_1)^2 a^\dagger_2 b^\dagger_2 + 2  ( \alpha \omega + \beta \gamma)  a^\dagger_1 b^\dagger_1 a^\dagger_2 b^\dagger_2 \Big] |\text{vac} \rangle  .
\label{spinorn2ex}
\end{align}
where we have defined $ a_m \equiv a_{m,0}, b_m \equiv a_{m,1} $ for notational simplicity.
Comparing coefficients, for most of the terms there is no problem in establishing a mapping between the spinor and spin coherent case.  For instance, looking at the term with coefficient $ \alpha \beta$, the bosonic state maps to
\begin{align}
& ( a^\dagger_1)^2 a^\dagger_2 b^\dagger_2 | \text{vac} \rangle \leftrightarrow | 00 \rangle | 01 \rangle + | 01 \rangle |00 \rangle  \nonumber \\
& = | 0 \rangle^{\otimes 2}_1 ( |01 \rangle_2 + | 10 \rangle_2 )
\end{align}
where in the last line we rearranged and labeled the qubits to better show which subsystem they lie on. We see that in this case the states are consistent with the mapping given in Appendix \ref{app:equiv} (see also Ref. \cite{byrnes2021quantum}).  We note we have not considered the proper normalization, and only discuss the identity of the states.  

Issues arise when considering the terms in the last lines of (\ref{scsn2ex}) and (\ref{spinorn2ex}).  Mapping the $ \alpha \omega $  and the $\beta \gamma $ terms we require two different states to both map to the same bosonic state
\begin{align}
a^\dagger_1 b^\dagger_1 a^\dagger_2 b^\dagger_2 | \text{vac} \rangle & \overset{?}{\leftrightarrow} | 00 \rangle | 11 \rangle + | 11 \rangle |00 \rangle \nonumber \\
a^\dagger_1 b^\dagger_1 a^\dagger_2 b^\dagger_2 | \text{vac} \rangle &  \overset{?}{\leftrightarrow} | 01 \rangle | 10 \rangle + | 10 \rangle | 01 \rangle  .
\end{align}
The states on the right hand side are distinct states, and it is problematic that distinct states do not exist in the spinor formulation.  Furthermore, both of the mappings above are inconsistent with the mapping of Appendix \ref{app:equiv}, since one would expect the mapping 
\begin{align}
a^\dagger_1 b^\dagger_1 a^\dagger_2 b^\dagger_2 | \text{vac} \rangle & \leftrightarrow \frac{1}{4} ( | 00 \rangle | 11 \rangle + | 11 \rangle |00 \rangle +  | 01 \rangle | 10 \rangle +  | 10 \rangle | 01 \rangle  )  \nonumber \\
& = \frac{1}{4} ( |01 \rangle_1 + | 10 \rangle_1 ) ( |01 \rangle_2 + | 10 \rangle_2 ) .
\end{align}
A consistent mapping is however possible if one assumes $ \alpha \omega = \beta \gamma $, which is satisfied for unentangled states.  

The above shows that a precise mapping between spin coherent state and spinor states is no longer possible with multipartite states, in the general case. The difference between the two classes of states originates from the different types of symmetry that they possess.  As suggested in (\ref{scsn2ex}), spin coherent states are symmetric under interchange of any two molecules.  This means one must interchange all the subsystems together.  Meanwhile, spinor states are symmetric under {\it local} interchange of the particles.  The spinor state consists of a superposition of a tensor product of Fock states (\ref{multifock}), which are symmetric under interchange of any two particles on the $ m$th subsystem.  In this way, the two classes of states are distinct.  Nevertheless, it is clear from their definitions that they both contain the same information of the underlying multipartite state $  \Psi $. Hence they can be considered different ways of storing the same quantum information, both in a highly duplicated way.

\section{The bipartite spinor state}
\label{sec:bipartite}

We now examine a specific example of a spinor state to gain more intuition about their properties. We consider the $ M = 2$ (bipartite) $ L = 2 $ (two-level) case for general $ N $ (duplication factor).

\subsection{State parameterization}

Let us consider a general bipartite two-level spinor state
\begin{align}
| \Psi \rangle \rangle = & \frac{1}{\sqrt{{\cal N}_\Psi }} 
 \left( \Psi_{00} a^\dagger_1 a^\dagger_2 + 
 \Psi_{01} a^\dagger_1 b^\dagger_2 + 
 \Psi_{10} b^\dagger_1 a^\dagger_2 
 + \Psi_{11} b^\dagger_1 b^\dagger_2 \right)^N \nonumber \\
 & \times   | \text{vac} \rangle  ,
\label{bipartitespinor}
\end{align}
where we again use the same definitions of the bosonic operators as in (\ref{spinorn2ex}).  Using the form preserving transformation of Sec. \ref{sec:formspinor}, we may write the spinor state in Schmidt form
\begin{align}
| \Psi \rangle \rangle = \frac{1}{\sqrt{ {\cal N}_\Psi }} V_1 V_2  \left(  \cos \chi a^\dagger_1 a^\dagger_2 + \sin \chi b^\dagger_1 b^\dagger_2 \right)^N | \text{vac} \rangle  ,
\label{schmidttwoqubit}
\end{align}
where for $ m \in \{1,2 \}$
\begin{align}
V_m = \exp(-i \bm{n}_m \cdot \bm{S}_m \theta_m/2) ,
\label{localspin}
\end{align}
and $ \bm{n}_m $ is a unit vector with 3 real components,  $ \bm{S}_m = (S^x_m, S^y_m, S^z_m) $.  The parameters $\bm{n}_m $ and $ \theta_m $ are the same parameters that would be chosen to put the unduplicated version of the state 
\begin{align}
\Psi_{00} |00 \rangle + \Psi_{01} |01 \rangle +\Psi_{10} |10 \rangle + \Psi_{11} | 11 \rangle
\end{align}
into Schmidt form.  

The normalization is best evaluated in Schmidt form.  Using (\ref{schmidttwoqubit}) the normalization factor is evaluated as
\begin{align}
{\cal N}_{\Psi} = & \langle \text{vac} |  \big(  \cos \chi a_1 a_2 + \sin \chi b_1 b_2 \big)^N \nonumber \\
& \times \big(  \cos \chi a^\dagger_1 a^\dagger_2 + \sin \chi b^\dagger_1 b^\dagger_2 \big)^N | \text{vac} \rangle 
\nonumber \\
& = (N!)^2 \sum_{k=0}^N \cos^{2k} \chi  \sin^{2N-2k} \chi \label{secondlastnorm} \\
& = \frac{ (N!)^2  (\cos^{2N+2} \chi - \sin^{2N+2} \chi)}{\cos 2\chi} ,
\end{align}
where in the first line we used $ V_m^\dagger V_m = I $ and in the second line we expanded the brackets and used the normalized Fock states (\ref{multifock}). We see explicitly the state-dependent aspect of the normalization factor. 

For the case $ \chi = \pi/4$, we separately evaluate
\begin{align}
{\cal N}_{\Psi} =   \frac{(N!)^2}{2^N} (N+1)
\end{align}
following from (\ref{secondlastnorm}).

\subsection{Hamiltonian}

In order to prepare the bipartite spinor state it is useful to know what Hamiltonian has (\ref{bipartitespinor}) as its ground state. We first start by deducing the Hamiltonian that has
\begin{align}
| \chi \rangle \rangle = \frac{1}{\sqrt{ {\cal N}_\Psi }}  \left(  \cos \chi a^\dagger_1 a^\dagger_2 + \sin \chi b^\dagger_1 b^\dagger_2 \right)^N | \text{vac} \rangle  
\label{schmidttwospinor}
\end{align}
as its ground state. It may be verified that the Hamiltonian
\begin{align}
H_0 = \sin 2 \chi ( S^y_1 S^y_2 - S^x_1 S^x_2) + \cos 2 \chi ( S^z_1 + S^z_2 ) - S^z_1 S^z_2
\label{twospinorham}
\end{align}
satisfies the eigenvalue equation
\begin{align}
H_0 | \chi \rangle \rangle = -N(N+2) | \chi \rangle \rangle 
\end{align}
for all $ N $. Hence the general spinor state (\ref{schmidttwoqubit}) may be created by finding the lowest energy state of 
\begin{align}
H = V_1 V_2 H_0 V_1^\dagger V_2^\dagger .  
\end{align}
The eigenvalue equation that this satisfies is
\begin{align}
H |\Psi \rangle \rangle = - N(N+2) | \Psi \rangle \rangle .
\end{align}

\subsection{Expectation values, covariances, and correlations}

\begin{figure}[t]
\includegraphics[width=\linewidth]{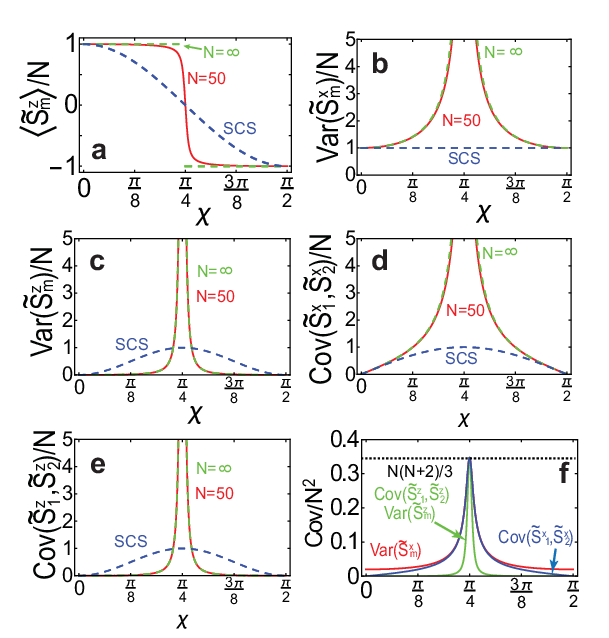}
\caption{Expectation values and covariances of spin operators for the two qubit spinor state (\ref{schmidttwoqubit}).  Spins are taken to be in the Schmidt basis. Solid lines correspond to exact values using the expressions given in Appendix \ref{app:expectation}, where $ N =50$ is used for all plots.  Dashed lines indicate approximations for large $ N $ or the values for the spin coherent states (SCS) as indicated. The remaining non-zero covariances are $ \text{Var} (\tilde{S}^y_m) = \text{Var} (\tilde{S}^x_m)$ and $ \text{Cov} (\tilde{S}^y_1 \tilde{S}^y_2) = -\text{Cov} (\tilde{S}^y_1 \tilde{S}^y_2) $.  (f) shows the same spinor variances and covariances as (b)-(e) but plotted on a larger scale to show their maximum values.  The dotted line shows the limiting value () for all curves.  
\label{fig2}  }
\end{figure}

Spin operators transform under the unitary operations (\ref{localspin}) in exactly the same way as for Pauli operators.  The spin transformation is 
\begin{align}
 V_m^\dagger S_m^i V_m = \sum_j O^{ij}_m S_m^j \equiv \tilde{S}_m^i
 \label{spintransforms}
\end{align}
where $ i,j \in \{x,y,z \} $ and $ O_m $ is the rotation matrix on the Bloch sphere associated with the unitary transformation (\ref{localspin}).  We indicate spin operators that are in the Schmidt basis with a tilde.  

While the local transformations give an equivalent transformation of the spin operators in comparison to the usual qubit  ($ N = 1$) Pauli spin operators, the remaining part of the wavefunction has a different behavior.  We therefore work in the Schmidt basis and evaluate the expectation values and covariances as given in Appendix \ref{app:expectation}.  The exact formula are given in Appendix \ref{app:expectation}, here we give some approximate formula that are valid in the limit of large $ N $.  

We will compare the expectation values and covariances of the spinor state to the corresponding spin coherent state
\begin{align}
| \Psi \rangle^{\otimes N} = V_1 V_2 \left( \cos \chi | 00 \rangle + \sin \chi | 11 \rangle \right)^{\otimes N} . 
\label{scsbipartite}
\end{align}
where the transformation to the Schmidt basis is the same as for the spinor case (\ref{localspin}).

\subsubsection{Single spin expectation values}

For the first order expectation values we have for $ m \in \{1,2 \} $:
\begin{align}
\langle \tilde{S}^x_m \rangle & = \langle \tilde{S}^y_m \rangle   = 0
\end{align}
which is the same as obtained for spin coherent states.  

For the diagonal spin operator in the Schdmit basis we have for large $ N $
\begin{align}
\langle \tilde{S}^z_m \rangle & \approx N \frac{ \cos 2 \chi }{ | \cos 2 \chi | }
=  N \text{sgn} ( \cos 2 \chi ) .
\label{szapprox}
\end{align}
In Fig. \ref{fig2}(a) we plot the normalized expectation value and also plot the limiting case for $ N \rightarrow \infty $.  We also compare it to the spin coherent state which in this case is $ \langle \tilde{S}^z_m \rangle = N \cos 2 \chi $. The spin coherent case simply $ N $ times the qubit expectation value, following (\ref{uniexp}).  Hence as $ N $ increases from $N =1 $, the original cosine function approaches a step function.   The origin of this sharper dependence are the additional factorials introduced in the Fock state definitions (\ref{multifock}) 
with the second subsystem.

\subsubsection{Covariances}

It is convenient to summarize the covariances in terms of a symmetrized covariance matrix defined with matrix elements as
\begin{align}
{\cal V}_{jk} = \frac{1}{2} \langle \{ \xi_j , \xi_k \} \rangle - \langle \xi_j \rangle \langle \xi_k \rangle
\label{covmatrix}
\end{align}
where $ \{C, D \} = CD + DC $ is the anticommutator and we take the operator set as 
\begin{align}
\xi = ( \tilde{S}^x_1, \tilde{S}^y_1,\tilde{S}^z_1,\tilde{S}^x_2, \tilde{S}^y_2,\tilde{S}^z_2) .
\label{operatorset}
\end{align}
In the limit of large $ N $ and away from the vicinity of $ \chi = \pi/4$, we approximately obtain (see Appendix \ref{app:expectation}) 
\begin{align}
& {\cal V} = \nonumber \\
&  \left(
\begin{array}{cccccc}
\frac{N}{| \cos 2x |} & 0 & 0 & \frac{N \sin 2 x }{| \cos 2x |} & 0 & 0 \\
0 & \frac{N}{| \cos 2x |} &  0 & 0 & -\frac{N\sin 2 x }{| \cos 2x |} &  0 \\
0 & 0 & \tan^2 2x  & 0 & 0 & \tan^2 2x \\
\frac{N\sin 2 x }{| \cos 2x |} & 0 & 0 & \frac{N}{| \cos 2x |} & 0 & 0 \\
0 & -\frac{ N \sin 2 x }{| \cos 2x |} &  0 & 0 & \frac{N}{| \cos 2x |} &  0 \\
0 & 0 & \tan^2 2x & 0 & 0 & \tan^2 2x
\end{array}
\right) .
\label{spinorcov}
\end{align}
This can be compared to the (exact) covariance matrix for the spin coherent state (\ref{scsbipartite}) which is 
\begin{align}
& {\cal V} = \nonumber \\
& N \left(
\begin{array}{cccccc}
1 & 0 & 0 & \sin 2 x  & 0 & 0 \\
0 & 1 &  0 & 0 & - \sin 2 x  &  0 \\
0 & 0 & \sin^2 2x & 0 & 0 & \sin^2 2x\\
 \sin 2 x & 0 & 0 & 1 & 0 & 0 \\
0 & - \sin 2 x &  0 & 0 & 1 &  0 \\
0 & 0 & \sin^2 2x & 0 & 0 & \sin^2 2x
\end{array}
\right) ,
\label{scscov}
\end{align}
which can be simply evaluated using (\ref{bipartiteexp}).  

We firstly see that all zero elements of the covariance matrix are in common between the spinor and spin coherent state versions.  This arises due to the fact that in an expansion in terms of Fock states, 
\begin{align}
| \chi \rangle \rangle = \frac{1}{\sqrt{{\cal N}_\Psi }}
\sum_k \cos^k \chi \sin^{N-k} \chi |k\rangle_1  |k \rangle_2  
\end{align}
and only operators that preserve the relative Fock number are non-zero.  Hence the only non-zero two-spin expectation values are $ \langle S_1^j S_2^j \rangle $ and $ \langle ( S_m^j )^2 \rangle $ for $ j \in \{ x,y, z \} $.

A comparison of the variances of local spin operators is shown in Fig. \ref{fig2}(b)(c).  For the variance of $ S^x_m $ and $ S^y_m$ the variance takes larger values than that of a spin coherent state.  As the state approaches the maximally entangled point $ \chi = \pi/4 $, the large $ N $ variance magnitude increases due to the factor of $ | \cos 2x |$ in the denominator in (\ref{spinorcov}).  For the variance of $ S^z_m $, for much of the domain of $ \chi$, it takes values less than that of the spin coherent state.  This is due to the fact that terms proportional to $ N $ and $ N^2 $ are suppressed and only the constant term survives (see Appendix \ref{app:expectation}).  Again near $ \chi = \pi/4$ the variance increases and overtakes the value for the spin coherent state. While the variances in (\ref{spinorcov}) appear to diverge, in fact for finite $ N $, all the variances approach a finite value
\begin{align}
\lim_{\chi \rightarrow \pi/4} \text{Var} ( \tilde{S}^j_m) = 
 \frac{N(N+2)}{3},  
\end{align}
for $ j \in \{x,y,z \}$.  This may be also seen in Fig. \ref{fig2}(f), where a larger range of the variances are plotted.  

The covariances show a similar dependence, these are plotted in Fig. \ref{fig2}(d)(e). First, we see the expected pattern of correlations for $ \text{Cov} (S^x_1, S^x_2), \text{Cov} (S^z_1, S^z_2)$ and anticorrelations for $ \text{Cov} (S^y_1, S^y_2) = -\text{Cov} (S^x_1, S^x_2)$.  The magnitude of the covariances $ \text{Cov} (S^x_1, S^x_2), \text{Cov} (S^y_1, S^y_2) $ exceed that of a spin coherent state while $ \text{Cov} (S^z_1, S^z_2) $ takes typically smaller values.  Near $ \chi =\pi/4$, the covariances take large values, but take a limiting value of 
\begin{align}
\lim_{\chi \rightarrow \pi/4}  \text{Cov} ( \tilde{S}^j_1 \tilde{S}^j_1)  = (-1)^{\delta_{jy}}
 \frac{N(N+2)}{3},  
\end{align}
for $ j \in \{x,y,z \}$ and $ \delta_{ij} $ is a Kronecker delta.  This is shown in Fig. \ref{fig2}(f).

\subsubsection{Correlations}

The different dependences of the covariances for the spinor states and spin coherent state suggest that quantitatively, these states have different correlations.  In fact, they are quite closely related as may be seen by looking at their correlations, which can be defined as
\begin{align}
\text{Corr} (C,D) = \frac{\text{Cov}(C,D)}{\sqrt{\text{Var}(C)\text{Var}(D)}} . 
\label{corrdef}
\end{align}
In effect, this quantity normalizes the covariance with the variance of the underlying variables themselves.  The large covariance as seen in Fig. \ref{fig2}(d)(e) may be understood as arising from the large variance of the original single spin variables (Fig. \ref{fig2}(b)(c)).  

Evaluating this for the spinor states we find
\begin{align}
\text{Corr} (\tilde{S}^x_1,\tilde{S}^x_1) & = \sin 2 \chi \nonumber \\
\text{Corr} (\tilde{S}^y_1,\tilde{S}^y_1) & = -\sin 2 \chi 
\nonumber \\
\text{Corr} (\tilde{S}^z_1,\tilde{S}^z_1) & = 1 .
\end{align}
Remarkably, these are {\it exact} relations valid for all $ N $, as may be found using the general  expressions in Appendix \ref{app:expectation}.  Furthermore, exactly the same expression are obtained for the spin coherent states, as may be easily verified from (\ref{scscov}).  In this sense, the spinor and spin coherent states have the same two-spin correlations.

In summary, spinor expectation values and covariances follow the same pattern of zero values.  For the non-zero expectation values they do not have the same type of dependence as the spin coherent state versions of the expectation values, although there is a resemblance in terms of sign and turning points. They however have {\it exactly} the same two-spin correlations as defined by (\ref{corrdef}). 
Local spin transformations work in exactly the same was as for qubits, due to the relations (\ref{spintransforms}).  Hence any difference between spin coherent state expectation values and the spinor expectation values will result from a different dependence of the non-zero expectation values and coavariances as shown in Fig. \ref{fig2}.

\subsection{EPR correlations}

Due to the same way that spinor states transform under local transformations to their unduplicated counterpart, they share similar entanglement properties.  This is best illustrated with the spinor version of the Bell state
\begin{align}
| \text{EPR}  \rangle \rangle =\frac{1}{N! \sqrt{N+1} }  \left( a_1^\dagger a_2^\dagger + b_1^\dagger b_2^\dagger \right)^N | \text{vac} \rangle  ,
\label{eprz}
\end{align}
which is the point $ \chi = \pi/4 $ in (\ref{schmidttwospinor}).
Making an expansion in the Fock basis we observe that this is a maximally entangled state
\begin{align}
| \text{EPR}  \rangle \rangle = \frac{1}{\sqrt{N+1}} \sum_{k=0}^N | k\rangle_1 | k \rangle_2  .  
\label{eprstatefock}
\end{align}
In Ref. \cite{kitzinger2020two} it was found that such states have a basis invariance property in Fock space.  Here we show the same result in a simpler way, using the spinor form.  

Making a change of basis to the $ x$-basis we have for $ m \in \{ 1,2 \}$
\begin{align}
a_m & = \frac{a_m^{x} + b_m^{x}}{\sqrt{2}} \nonumber \\
b_m & = \frac{a_m^{x} - b_m^{x}}{\sqrt{2}}
\end{align}
such that we may rewrite
\begin{align}
| \text{EPR}  \rangle \rangle =\frac{1}{N! \sqrt{N+1} }  \left( {a_1^{x}}^\dagger {a_2^{x}}^\dagger + {b_1^x}^\dagger {b_2^x}^\dagger \right)^N | \text{vac} \rangle  .
\label{eprx}
\end{align}
In the $y$-basis the operators transform as
\begin{align}
a_m & = \frac{a_m^{y} + i b_m^{y}}{\sqrt{2}} \nonumber \\
b_m & = \frac{ i a_m^{y} + b_m^{y}}{\sqrt{2}}
\end{align}
\begin{align}
| \text{EPR}  \rangle \rangle =\frac{(-i)^N}{N! \sqrt{N+1} }  \left( {a_1^{y}}^\dagger {b_2^{y}}^\dagger + {b_1^y}^\dagger {a_2^y}^\dagger \right)^N | \text{vac} \rangle  .
\label{epry}
\end{align}

The states (\ref{eprz}), (\ref{eprx}), and (\ref{epry}) show the same pattern EPR correlations and anti-correlations. Quantitatively, we have
\begin{align}
\text{Var} ( S^z_1 - S^z_2) = \text{Var} ( S^x_1 - S^x_2) = \text{Var} ( S^y_1 + S^y_2) = 0, 
\label{variances}
\end{align}
which may be evaluated by expanding in the Fock basis.  For $ N $ large, this can be described as a macroscopic EPR state.

\subsection{Entanglement}

\begin{figure}[t]
\includegraphics[width=\linewidth]{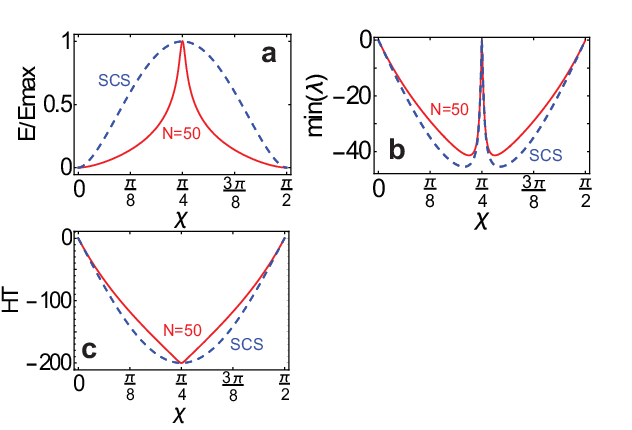}
\caption{Entanglement in the bipartite spinor state (\ref{bipartitespinor}) and spin coherent state (\ref{scsbipartite}). (a) Von Neumann entropy normalized to the maximum entanglement $ E_{\max} = \log_2 N+1 $ for the spinor state and $  E_{\max} = \log_2 2^N = N $ for the spin coherent state.  (b) Minimum eigenvalue of the left hand side of (\ref{covmatrixcrit}) and (c) left hand side of the Hoffman-Takeuchi (HT) inequality (\ref{hoffmantakeuchi}).  In (b) and (c) negative values indicate the presence of entanglement.  In all cases $ N = 50$.  
\label{fig5}  }
\end{figure}

The EPR state (\ref{eprstatefock}) where $ \chi = \pi/4$ is a maximally entangled state.  Quantifying the entanglement for arbitrary $ \chi $ can be done using entanglement measures such as the von Neumann entropy, given by
\begin{align}
E= - \text{Tr} (\rho \log_2 \rho ) = - \sum_k \lambda_k \log_2  \lambda_k 
\end{align}
where $ \rho = \text{Tr}_2 | \Psi \rangle \rangle \langle \langle \Psi | $ and $ \lambda_k $ are the eigenvalues of $ \rho$.  In this case we have $ \lambda_k = \cos^{2k} \chi \sin^{2N-2k} \chi/{\cal N}_\Psi  $ and the entropy of plotted in Fig. \ref{fig5}(a).  We see that the entropy gives the maximum value $ E_{\max} = \log_2 (N+1) $ as expected. For a spin coherent state the entanglement is also a maximum at $ \chi = \pi/4 $ but it reaches a larger value $ E_{\max} = \log_2 2^N = N $ due to the larger Hilbert space that is available.  

For continuous variables quantum optics, the covariance matrix (\ref{covmatrix}) plays a central role in characterizing states.  In particular, it can be used to detect entanglement between modes using Simon's criterion \cite{simon2000peres}.  A generalization of this to arbitrary operators was previously performed, where it was shown that for separable states \cite{tripathi2020covariance}
\begin{align}
\text{PT} ({\cal V}) + \frac{i}{2} \text{PT} (\Omega) \ge 0 ,
\label{covmatrixcrit}
\end{align}
where $ \text{PT} $ performs the partial transpose operation on the operators involved in the expectation values. Here, $ \Omega $ is the commutation matrix defined with matrix elements $ \Omega_{jk} = -i \langle [\xi_j, \xi_k ] \rangle $. For the operators defined in (\ref{operatorset}), the commutation matrix is
\begin{align}
& \Omega =  2 \left(
\begin{array}{cccccc}
0 & \langle \tilde{S}^z_1 \rangle & - \langle \tilde{S}^y_1 \rangle  & 0  & 0 & 0 \\
- \langle \tilde{S}^z_1 \rangle & 0 & \langle \tilde{S}^x_1 \rangle & 0  & 0 & 0 \\
\langle \tilde{S}^y_1 \rangle & -\langle \tilde{S}^x_1 \rangle & 0 & 0  & 0 & 0 \\
0 & 0 & 0 & 0 & \langle \tilde{S}^z_2 \rangle & - \langle \tilde{S}^y_2 \rangle  \\
0 & 0 & 0 & - \langle \tilde{S}^z_2 \rangle & 0 & \langle \tilde{S}^x_2 \rangle \\
0 & 0 & 0 & \langle \tilde{S}^y_2 \rangle & -\langle \tilde{S}^x_2 \rangle & 0
\end{array}
\right) .
\label{scsomega}
\end{align}
The partial transpose version of $ {\cal V} $ involves changing the sign of $ \tilde{S}^y_2 $ \cite{tripathi2020covariance} and corresponds to removing the minus sign (and thereby making them positive) in the elements of (\ref{spinorcov}) and (\ref{scscov}).  Meanwhile, for the commutation matrix, since $ \langle \tilde{S}^y_2 \rangle = 0 $ in our case, $ \text{PT} (\Omega) = \Omega $.  Evaluating the minimum eigenvalue of the left hand side of (\ref{covmatrixcrit}) is shown in Fig. \ref{fig5}(b). 
For the spinor covariance matrix, we use the exact expressions as given in Appendix \ref{app:expectation}, not the approximate matrix (\ref{spinorcov}). 
We see that entanglement is successfully detected except for the maximally entangled point $ \chi = \pi/4$.  This point fails due to $ \Omega = 0 $ at this point, where the approach fails. However, for all remaining points entanglement is successfully detected. 

To better handle entanglement detection in the vicinity of  $ \chi = \pi/4$, other approaches are available, such as those discussed in Refs. \cite{jing2019split,gao2023optical,guhne2009entanglement}. Among these, one of the best performing criteria is the Hoffman-Takeuchi inequality \cite{hofmann2003violation}. 
This criterion is expecially appropriate in our case since the Holstein-Primakoff approximation is not used.  This criterion states that for separable states 
\begin{align}
\text{Var} ( S^x_1 - S^x_2) +  \text{Var} ( S^y_1 + S^y_2)  +\text{Var} ( S^z_1 - S^z_2) - 4 N  \ge  0 .
\label{hoffmantakeuchi}
\end{align}
Fig. \ref{fig5}(c) shows a plot of the left hand side of (\ref{hoffmantakeuchi}) which shows that entanglement is detected in the full range, and reaches the maximum violation at $ \chi = \pi/4 $.  Interestingly, despite the different amount of entanglement between the spinor and spin coherent states, the level of the violation for the two states are the same at $ \chi = \pi/4$.

\subsection{Wigner functions}

We next visualize the bipartite spinor state using the spin Wigner function \cite{dowling1994wigner,byrnes2021quantum}.  Generalizing the spin Wigner function to the multipartite case with equal dimensions on each subsystem we have
\begin{align}
& W( \theta_1, \phi_1, \dots, \theta_M, \phi_M ) = \sum_{L_1=0}^{2j} \dots \sum_{L_M=0}^{2j}  \sum_{m_1=-j}^j \dots \sum_{m_M=-j}^j \nonumber \\
& \times \sum_{m_1'=-j}^j \dots \sum_{m_M' =-j}^j  \rho_{\vec{m} \vec{m}'}  (-1)^{Mj + \sum_{n=1}^M m_n' } \nonumber \\
& \times \prod_{n=1}^M 
\langle j, m_n; j, -m_n' | L_n, m_n- m_n' \rangle  Y_{L_n, m_n- m_n'} (\theta_n, \phi_n) ,
\end{align}
where we have changed notation from the Fock state notation (\ref{multifock}) to angular momentum, which have an equivalence according to the Jordan-Schwinger representation as \cite{byrnes2021quantum}
\begin{align}
|j, m \rangle = | k = j+m \rangle = \frac{(a^\dagger)^{j+m} (b^\dagger)^{j-m}}{\sqrt{(j+m)! (j-m)!}}.  
\end{align}
Here $ j = N/2 $ is the total angular momentum quantum number and $ m \in [-j, j] $ is the $z$-projection quantum number.  The matrix elements $ \langle j, m; j, m' | L, M \rangle$ are Clebsch-Gordan coefficients which are non-zero only if $ M = m+ m'$, and $Y_{L,M} (\theta, \phi)$ are the spherical harmonics. The matrix elements are defined as 
\begin{align}
 \rho_{\vec{m} \vec{m}'} = \langle j, m_1 | \dots \langle j, m_M | \rho | j, m_1' \rangle \dots | j, m_M' \rangle. 
\end{align}
and we denoted $ \vec{m} = (m_1, \dots, m_M ) $. 

The Wigner function has the property that a local spin unitary transformation (\ref{localspin}) rotates the spherical distributions in $ (\theta_m, \phi_m) $ on the sphere.  For our bipartite spinor state (\ref{schmidttwoqubit}), it will therefore be sufficient to consider the state in the Schmidt basis (\ref{schmidttwospinor}).  The effect of the remaining unitary rotations in (\ref{schmidttwoqubit}) can be deduced by rotations on the Bloch sphere.  

\begin{figure}[t]
\includegraphics[width=\linewidth]{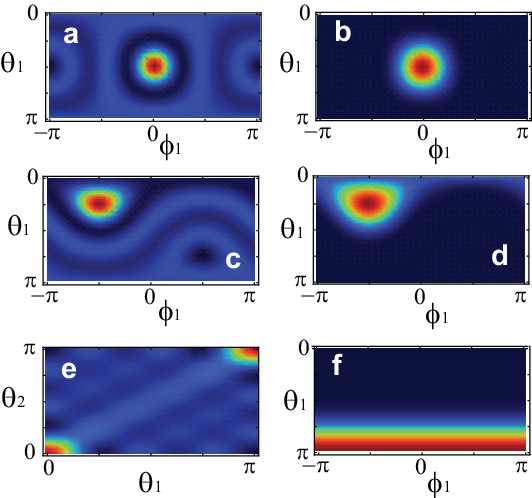}
\caption{Wigner functions for bipartite spinor state.  Bipartite Wigner functions for fixed (a) $ \theta_2 = \pi/2, \phi_2 = 0 $; (c)  $ \theta_2 = \pi/4, \phi_2 = \pi/2 $.  Unipartite Wigner functions for the state (\ref{projected}) with parameters (b) $ \theta_2 = \pi/2, \phi_2 = 0 $; (d)  $ \theta_2 = \pi/4, \phi_2 = \pi/2 $. (e) The 
marginal Wigner function (\ref{marginalwigner}).  (f) Unipartite Wigner function for (\ref{tracedoutrho}) with $ \chi = \pi/8$.   
\label{fig4}  }
\end{figure}

Figure \ref{fig4}(a)(c) show the bipartite Wigner functions for (\ref{schmidttwospinor}), which is a function of four parameters $ \theta_1, \phi_1,  \theta_2, \phi_2$. In order to visualize this distribution, we choose fixed values of $ \theta_2, \phi_2 $ and plot the remaining variables.  We observe that the the Wigner distributions are similar to spin coherent states that are centered around $ \theta_1 = \theta_2 $ and $ \phi_1 = - \phi_2 $.  To understand this relationship, we calculate the state after projecting out the second ensemble with a spin coherent state for the EPR state ($\chi = \pi/4 $)
\begin{align}
\frac{\langle \langle \theta_2 , \phi_2  | \text{EPR} \rangle \rangle}{\sqrt{ \langle \langle \text{EPR}  |   \theta_2 , \phi_2  \rangle \rangle \langle \langle  \theta_2 , \phi_2  | \text{EPR} \rangle \rangle}} = | \theta_2, -\phi_2 \rangle \rangle_1 
\label{projected}
\end{align}
which gives a state on subsystem 1.  Here we used the fact that the EPR state can be written \cite{kitzinger2020two}
\begin{align}
 | \text{EPR} \rangle \rangle = 
 \frac{1}{\sqrt{N+1}} \sum_{k=0}^N | k \rangle^{(\theta, \phi)} 
 | k \rangle^{(\theta, -\phi)} 
\end{align}
where 
\begin{align}
| k \rangle^{(\theta, \phi)}  = e^{-i S^z \phi /2} e^{-i S^y \theta /2} | k \rangle 
\end{align}
are Fock states in a rotated basis and we used the fact that $ |k = N \rangle^{(\theta, \phi)} = | \theta, \phi \rangle \rangle $. Figure \ref{fig4}(b)(d) show the unipartite Wigner functions for the state (\ref{projected}).  We see an obvious resemblance to Fig. \ref{fig4}(a)(c).  The Wigner functions hence serve as a visualization of the correlations that exist between the two subsystems.  

Figure \ref{fig4}(e) shows the marginal Wigner function for the variables $ \theta_1, \theta_2 $, defined as 
\begin{align}
W(\theta_1, \theta_2) = \int_0^{2 \pi} \int_0^{2 \pi} d \phi_1 d \phi_2 W(\theta_1, \phi_1, \theta_2, \phi_2) .
\label{marginalwigner}
\end{align}
We again see the correlation between the $ \theta_1 $ and $ \theta_2 $ variables.  As expected the main correlations appear along the diagonal $ \theta_1 = \theta_2 $ due to the presence of EPR correlations.  There is interestingly a higher concentration at the poles $ \theta=0, \pi$, which we attribute to a similar effect to that seen in Fig. \ref{fig2}(a) where the distribution tends to concentrate at the poles.  
Finally, in Fig. \ref{fig4}(f) we show the Wigner function for the state where subsystem 2 is traced out
\begin{align}
\rho_1 & = \text{Tr}_2 | \chi \rangle \rangle \langle \langle \chi | \nonumber \\
& = \frac{1}{N_\Psi} \sum_{k=0}^N \cos^{2k} \chi   \sin^{2N-2k} \chi  |k\rangle \langle k | . 
\label{tracedoutrho}
\end{align}
For the maximally entangled state  $\chi = \pi/4 $, the Wigner functions are completely uniform in $ \theta_1, \phi_1 $.  We therefore plot a partially entangled state $ \chi = \pi/8$, which shows a distribution reminiscent of a thermal state, which is featureless in the $ \phi$ direction, but is exponentially distributed in the $ \theta $ direction. 

The Wigner function shows features that are analogous to two-mode squeezed states in optical systems \cite{braunstein2005quantum}. For two-mode squeezed states, correlations are seen for between quadratures along the lines $ x_1 = x_2 $ and $ p_1 = - p_2 $. Here they are distributed on the angular variables on the Bloch sphere for each subsystem.  We emphasize that these are these correlations go beyond the Holstein-Primakoff mapping that is typically performed on atomic systems.  For the maximally entangled state $ \chi = \pi/4$, there is no single spin direction that is polarized, and (\ref{tracedoutrho}) gives a completely mixed state. With the exception of $ \chi $ in the vicinity of $ 0, \pi/2 $, the spin operators cannot be approximated by quadratures, and the Holstein-Primakoff approximation breaks down.

\section{Errors in quantum information storage}
\label{sec:error}

In this section, we discuss the potential application of multipartite spin coherent states (\ref{multiscs}) and spinor states (\ref{multiscsbosonic}) as a means of storing quantum information.  From the general form of these two classes of states it is obvious that both classes of states are capable realizing a quantum register for a  quantum computer --- in both cases they are simply $ N $-fold duplicates of a quantum register state.  
The difference between the two classes of states is the type of symmetry they obey.  The spinor states are symmetric under local particle interchange, while spin coherent states are symmetric under multipartite interchange (i.e. interchange of the molecules in Fig. \ref{fig1}(a)).  

An important issue is how well quantum information can be stored in such states in the presence of errors.  To qualify as a good way to store quantum information, the states must not be excessively sensitive to errors, and must be readily accessible under measurement.  To give a contrasting example, a particularly poor way of storing quantum information would be to use Schrodinger cat state qubits,
i.e. $ \alpha |0,0 \rangle^{\otimes N} + \beta | \pi, 0 \rangle^{\otimes N} $ in the notation of (\ref{qubitscs}) and $ \alpha, \beta$ are complex coefficients.  Such states are known to decohere extremely quickly and any quantum information would be lost easily. 
In addition, to read out the coherence between the qubits (i.e. an $ x$-basis type measurement), high-order spin-changing interactions must be measured, which may be experimentally challenging. Here we discuss the potential advantages as compared to the unduplicated $ N =1 $ case when stored as spin coherent and spinor states.

\subsection{Spin coherent states}

Let us consider single particle errors given by Kraus operators $ E_{n}^{(l)}$, which act on the $ n $th molecule, and $ l $ labels the error type.  These satisfy $ \sum_l {E_{n}^{(l)}}^\dagger E_{n}^{(l)} = I $.  Then a particular error instance on the spin coherent state is 
\begin{align}
{\cal E}_{\vec{l}} = \prod_{n=1}^N E_{n}^{(l_{n})} ,
\end{align}
where $ \vec{l} =(l_{1}, \dots, l_{n}, \dots , l_{N}) $. We consider such errors as the most likely form of errors to occur, since typically single particle errors are most probable.  In fact, errors of the form $ E_n^{(l)} $ could also include multiparticle interactions between the same molecule, hence our model is more general than a single particle error model.  We furthermore assume that the error types are identical across all the molecules, i.e. $ E_{n}^{(l)} = E_{n'}^{(l)}$. Again this is physically reasonable if the ensemble is physically located in the same region.  

For this type of error, the state after the error is 
\begin{align}
\sum_{ \vec{l} }  {\cal E}_{\vec{l}} | \Psi \rangle^{\otimes N}  \langle \Psi |^{\otimes N}  {\cal E}_{\vec{l}}^\dagger  = \rho^{\otimes N} ,
\label{productrho}
\end{align}
where $ \rho = \sum_l E_n^{(l)} | \Psi \rangle \langle \Psi | {E_n^{(l)}}^\dagger $ is the density matrix of the $ n$th molecule.  For symmetric observables of the form (\ref{symmetricop}), the relations as that discussed in Sec. \ref{sec:expcov} still hold for such a state.  For example, 
\begin{align}
\langle C \rangle_{\rho^{\otimes N}} & = N \langle c \rangle_{\rho} \nonumber \\
\text{Var}_{\rho^{\otimes N}} (C) & = N \text{Var}_{\rho} (c )
\label{expectationvalueerror}
\end{align}
where we have made explicit what state the expectation values are being taken with a subscript. The relations (\ref{expectationvalueerror}) show that for symmetric observables, spin coherent states have an equivalent performance to the original states. Namely, if one uses normalized variables such as $ \langle C \rangle/N $ and $ \text{Var}(C)/N $, there is no longer any $ N $-dependence and exactly the same averages are obtained as the microscopic versions.  This is as expected, since spin coherent states are simply product states and are independent.  

For symmetric observables, (\ref{expectationvalueerror}) shows that the use of spin coherent states is no better but also no worse than the unduplicated case in the case of single particle decoherence. There are nevertheless some aspects which make them beneficial from a quantum information point of view.  The first is that the normalized noise of such symmetric observables have a scaling as
\begin{align}
\frac{\sqrt{\text{Var} (C)}}{\langle C \rangle } = \frac{1}{\sqrt{N} } \frac{\sqrt{\text{Var} (c ) }}{\langle c \rangle} . 
\label{signaltonoise}
\end{align}
Hence compared to the original microscopic versions, the noise of such observables is reduced by a factor of $ \sqrt{N} $. Hence a much larger signal-to-noise ratio is obtained with spin coherent states.  Of course, we note that this is the same statistical scaling as would be obtained from $ N $ runs of a quantum computer.  As discussed further in the next section, the advantage is present when spin coherent states could be generated in a single run of the experiment, such that it would not require the $N$-fold additional time resources.   

Another potential benefit is in a digital error correction scenario. Consider a situation where the information that is stored is classical binary digital information.  Reading out such a register is a common task in several important quantum algorithms.  For example, in Grover's algorithm, the final state readout of the register is the search target state in the $ z $-basis.  Another example is adiabatic quantum computing, where the quantum register contains the ground state configuration of the Ising Hamiltonian in the $ z$-basis.  Yet another example is the readout of the quantum phase estimation register, where the register shows the binary representation of the phase, also in the $ z$-basis. The common element of all these algorithms is that partway during the execution of the quantum algorithm, the quantum register is in a superposition state, but at the end of the algorithm the state is in a $z$-basis state of the form  $ \bigotimes_{m=1}^M | \sigma_m \rangle $ where $ \sigma_m \in  \{ 0, 1 \} $.  

In the spin coherent state case, in this scenario, the quantum register after the completion of the quantum algorithm is
\begin{align}
  \bigotimes_{m=1}^M | \sigma_m \rangle^{\otimes N}
\end{align}
in the ideal error-free case. In the presence of independent errors, this state is modified to 
\begin{align}
\bigotimes_{m=1}^M (p_m^{(0)} | 0 \rangle \langle 0 | + p_m^{(1)} | 1 \rangle \langle 1 | )^{\otimes N}
\label{errorensemblemaj}
\end{align}
where $ p_m^{(0)} + p_m^{(1)} =1 $.  We have assumed bit flip errors which is the only type of error that is relevant in this case (phase flip errors do not affect a $ z $-measurement).  In this case, we may take advantage of the duplicity of the spin coherent state.  For a particular ensemble labeled by $ m $, the probability of obtaining $ k $ bits in the state $ | 0 \rangle $ and $ N-k $ in the state $ | 1 \rangle $ is
\begin{align}
P_m(k) = \binom{N}{k} (1- \epsilon )^k \epsilon^{N-k}  ,
\end{align}
where we have taken $ \sigma_m = 0 $ without loss of generality and the error probability $ \epsilon = p_m^{(1)}$.  After a measurement of (\ref{errorensemblemaj}) is made, a majority vote is taken to obtain the final result.  The observable in this case is $ \text{sgn} (S^z_m) $ hence is not a symmetric observable as considered before.  

The logical error probability is then given by the error that more than half the bits in the ensemble are in the state $ | 1 \rangle $, given by
\begin{align}
\epsilon_L = \sum_{k=0}^{N/2} P_m(k) \approx \frac{[4 (1-\epsilon) \epsilon]^{N/2} }{\sqrt{N}} ,
\label{m1logicalerror}
\end{align}
where the approximation is valid for $ \epsilon \ll 1/2 $. We see that now there is an exponential suppression of logical errors in this situation.  
Figure \ref{fig6} shows the strong logical error suppression effect and the performance of the approximation.  This is possible because of the a priori knowledge that the state should be either in the state $ | 0 \rangle $ or $ |1 \rangle $.

\subsection{Spinor states}

For spinor states, we expect a similar behavior in the presence of errors, due to the similar duplication that is present in these states. Although it more difficult to make the same general statements that were made for spin coherent states, we show some specific examples which illustrate that analogous behavior is present.

First let us consider the effect of single particle errors on spin expectation values.  
Considering the case of bosonic loss, the Kraus operators are \cite{nielsen2002quantum}
\begin{align}
E^{(l)}_a 
= \sqrt{\frac{(1-\gamma)^l}{l!}} \sqrt{\gamma}^{a^\dagger a } a^l 
\end{align}
where $ 1- \gamma $ is the loss probability.  Now let us evaluate the Kraus operators by applying them on spin observables $ S^j $ for $ j \in \{x,y,z\}$, rather than on the state.  For collective spin operators, one may evaluate \cite{kitzinger2020two}
\begin{align}
    \sum_{l,l'=0}^\infty {E_a^{(l)}}^\dagger {E_b^{(l')}}^\dagger S^j E_b^{(l')} E_a^{(l)} = \gamma S^j ,
\end{align}
where we have assumed that the loss probability is the same for both $ a $ and $ b $ atoms.  We see that loss results in a $N$-independent factor multiplying the spin operators. This is again similar to the result (\ref{expectationvalueerror}) where the decoherence effect is equivalent to the microscopic case.  In this case, the factor of $ \gamma $ may in fact be accounted for by normalizing the spins to the number operator $ \hat{N} = a^\dagger a + b^\dagger b$ which obeys
\begin{align}
    \sum_{l,l'=0}^\infty {E_a^{(l)}}^\dagger {E_b^{(l')}}^\dagger \hat{N} E_b^{(l')} E_a^{(l)} = \gamma \hat{N} .
\end{align}
Hence measuring $ \langle S^j \rangle/ \langle \hat{N} \rangle $ removes the effect of atom loss for spin observables. Covariances can be handled in a similar way.  

The results for dephasing are similar.  The Kraus operator for dephasing in the $ z $-basis reads \cite{leviant2022quantum}
\begin{align}
E^{(l)}_{n} = \sqrt{\frac{\kappa^l}{l!}} e^{-\kappa n^2 /2} n^k 
\end{align}
where $ n = a^\dagger a$ and $ \kappa $ is the dephasing strength. In this case 
\begin{align}
\sum_{l=0}^\infty {E_n^{(l)}}^\dagger  S^x  E_n^{(l)} & = e^{-\kappa/2 } S^x  \nonumber \\
\sum_{l=0}^\infty {E_n^{(l)}}^\dagger  S^y  E_n^{(l)} & = e^{-\kappa/2 } S^y  \nonumber \\
\sum_{l=0}^\infty {E_n^{(l)}}^\dagger  S^z  E_n^{(l)} & =  S^z ,
\end{align}
and again we see a $N$-independent renormalization of the operators that are orthogonal to the $ z $-direction. This is similar to the result of (\ref{expectationvalueerror}) where the decoherence reduces to be the same as the microscopic version. Dephasing in other bases can be deduced by a simple basis change and give similar results. 
We note that the moderate effect of dephasing on these operators is thanks to the low order product of bosonic operators in $ S^j$.  Such observables only change the Fock state number by one unit at most, which gives rise to the factor of $ e^{-\kappa/2 }$.  States such as Schrodinger cat states are more severely affected by decoherence, where highly off-diagonal matrix elements are quickly degraded.  However, since in spin coherent states generally low order spin correlations are used to encode the quantum information, the effect is not as severe.  

For the signal-to-noise enhancement effect analogous to (\ref{signaltonoise}) for spinor states, we must again look at specific cases as it is difficult to make universal statements.  For unipartite spinor states, we have a mathematical equivalence to spin coherent states, hence we again have the same relation for the normalized noise of spin expectations
\begin{align}
\frac{\sqrt{\text{Var}(S^j)}}{\langle S^j \rangle }= \frac{1}{\sqrt{N}} \frac{\sqrt{\text{Var}(\sigma^j)}}{\langle \sigma^j \rangle } . 
\end{align}
Hence for unipartite spinor states there is a signal-to-noise enhancement with the same scaling as spin coherent states.  For the bipartite spinor state (\ref{schmidttwoqubit}), the only non-zero single spin expectation value is $ \tilde{S}^z_m$ and the normalized noise is
\begin{align}
\frac{\sqrt{\text{Var}(\tilde{S}^z_m)}}{\langle \tilde{S}^z_m \rangle } \approx \frac{\tan 2 \chi }{N} 
\label{spinorsignnoise}
\end{align}
which is valid for $ N \gg 1 $. So in this case, the signal-to-noise scaling in terms of $ N $ is in fact better than the spin coherent state case.  It does however become worse in the vicinity of the maximally entangled point $ \chi = \pi/4 $, where it is more difficult to read off expectation values than the spin coherent states.     

Finally, regarding the digital error correction scenario, the same results as the spin coherent states hold again due to the equivalence for unipartite states. Namely, as long as the error-free final state after the quantum evolution is of the form
\begin{align}
\bigotimes_{m=1}^M | \pi \sigma_m , 0 \rangle \rangle_m  
\label{idealfinalreg}
\end{align}
with $ \sigma_m = \{0, 1 \} $, the same conclusions to the spin coherent state case can be made due to the equivalence of unipartite spin coherent state and spinor states. 

As another error correcting scenario, consider a situation where due to gate errors, the states are not exactly (\ref{idealfinalreg}) but have a slight imperfection in terms of either an over- or 
under-rotation such that the state is instead $ | \theta, \phi \rangle \rangle $.  In this case we obtain the same result as (\ref{m1logicalerror}) with $ \epsilon = \sin^2 \frac{\theta}{2} $, which is the single particle error probability, assuming without loss of generality $ \sigma_m = 0 $. 

\begin{figure}[t]
\includegraphics[width=\linewidth]{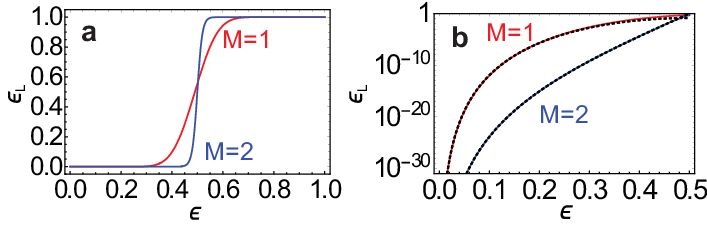}
\caption{Digital error correction with spin coherent states and spinor states.  The logical error probability $ \epsilon_L $ versus microscopic error probability $ \epsilon $. (a)  The exact expression single ensembles case $ M = 1$ given by (\ref{m1logicalerror}) and the two ensemble case $ M = 2 $ given by (\ref{m2logicalerror}).  (b) shows the same as (a) but on a semilog plot and showing the approximation in (\ref{m1logicalerror}) and (\ref{m2logicalerror}) as dotted lines. $ N=50 $ is used for all calculations. 
\label{fig6}  }
\end{figure}

For residual partial entanglement that is present in the final spinor state, consider a state of the form (\ref{schmidttwospinor}) where ideally $ \chi = 0 $, but there is some error $ \epsilon = \sin^2 \chi $ such that with some probability the $ \sigma_m = 1$ case is obtained.  This may occur in a quantum algorithm where the intermediate states are entangled, and due to gate errors the final states is not quite in the form (\ref{idealfinalreg}).  The logical error probability is then
\begin{align}
\epsilon_L = \frac{1}{{\cal N}_\Psi} \sum_{k=0}^{N/2} (1-\epsilon)^k \epsilon^{N-k} \approx \left( \frac{\epsilon}{1-\epsilon} \right)^{N/2}
\label{m2logicalerror}
\end{align}
As seen in Fig. \ref{fig6}, there is an even stronger logical error suppression effect when measuring $ \text{sgn} (\tilde{S}^z_m)  $ observable than the single ensemble case.  Such a error suppression technique was demonstrated explicitly in the context of adiabatic quantum computing with spinor states in Ref. \cite{mohseni2021error}.

\section{Experimental preparation}
\label{sec:preparation}

The primary purpose of this paper is to discuss multipartite extensions of spin coherent and spinor states and analyze their properties from a fundamental perspective.  We now discuss how such states may be realized experimentally. 

Unipartite spin coherent states can be realized in principle in numerous systems where there are a large number of duplicate controllable quantum systems.  A particularly suitable system is an atomic gas ensemble, where there are a large number of atoms of the same species and there is fixed particle number.  Such atomic ensembles may be trapped by either optical or magnetic traps, or be physically confined in a glass cell.  The levels $ |l \rangle $ appearing in (\ref{scsdef}) then refer to the internal states of the atoms. Typically the hyperfine ground states are used which have long coherence times. The motional degrees of freedom decouple to the internal degrees of freedom of the atoms, hence may be ignored \cite{hammerer2010quantum}.   Using optical pumping, the internal state of the atoms are polarized in a particular spin direction.  Once a polarized state is obtained, coherent unitary rotations, realized by optical or microwave/radio frequency radiation is applied in order to produce a more general spin coherent state. At the unipartite level, spin coherent states are routinely produced in atomic ensembles \cite{hammerer2010quantum,wieman1999atom,phillips1998nobel}. 

Unipartite spinor states are most directly realized in systems where there are degenerate bosons, such as in an atomic Bose-Einstein condensate \cite{bohi2009coherent,reichel2011atom,abdelrahman2014coherent}. In (\ref{scsdefboson}), the bosonic operators $ a_l$ annihilate an atom with internal state label $ l $.  Typically for a BEC, the condensation process occurs with respect to a particular internal spin state, hence no optical pumping is necessary.  Using the similar coherent manipulation methods as for thermal atomic gases, a spinor state of the desired form is produced.  We note that due to the mathematical equivalence of unipartite spin coherent states and spinor states, under the mapping in Appendix \ref{app:equiv}, in fact atomic ensembles can also realize spinor states. There is however a mathematical mapping required to connect them, since spinor states involve degenerate identical bosons, which are not strictly speaking present in a thermal atomic gas.  

For the multipartite spin coherent states, the natural extension would be for a molecular gas ensemble. This would be a close realization of the state that is pictured in Fig. \ref{fig1}(a), where each duplicate in the spin coherent state is literally a molecule.  In order to control the system, again, one would require an optical pumping scheme to polarize the state of all the molecules to the same state, and from there, coherent manipulation of the molecular states which illuminates all molecules would prepare the multiparticle spin coherent state.  Such a situation is reminiscent of Nuclear Magnetic Resonance (NMR), where control pulses are used to manipulate the states of molecules. The challenge in the case of liquid state NMR has been to realize a high fidelity NMR system that is challenging under present technology \cite{ladd2010quantum}.  Additionally, in both cases it is difficult to scale since as the system grows larger with $ M $, there are an exponentially larger number of states that must be discerned within each molecule. The problem arises due to the control fields hitting all subsystems of the molecules together, so that the molecular wavefunction is controlled as a whole.  This is in contrast to modern quantum computing architectures, where quantum control is performed by combining gates on subsystems, which is a more scalable approach.  

Another approach to prepare multipartite spin coherent states would be to simply have $N$ quantum computers which all prepare the same quantum state. This clearly would trivially prepare the state (\ref{multiscs}) since it is simply a product state of $ N $ quantum registers.  However, this is very expensive from a resource perspective, as individual microscopic control of each of the molecules is required.  For a molecular gas, there is little overhead in increasing the size of the ensemble $N $, as it merely involves obtaining a larger sample of the gas.  For the quantum control, the same control pulses illuminate the entire ensemble, which again does not require additional resources.  Hence the advantages as described in Sec. \ref{sec:error} are only really beneficial unless larger ensembles can be implemented without resources that scale as $ N $.  

For multipartite spinor states, the natural realization are multiple atomic ensembles or BECs that are entangled together. In Fig. \ref{fig1}(c), each of the local subsystems would be an atomic ensemble consisting of $ L $ internal states. 
The most direct implementation would be to form multiple BECs, for example on an atom chip, and entangle them together \cite{abdelrahman2014coherent}.
Using the equivalence in Appendix \ref{app:equiv} between atomic ensembles and degenerate bosons, it is also possible to use multiple atomic ensembles and entangle them together.  Multiple atomic glass cells have been entangled together \cite{julsgaard2001experimental,pu2017experimental,pu2018experimental} in the case of thermal atomic ensembles.  The most versatile way to perform the entangling operations are quantum nondemolition (QND) measurements, which can be realized by sending a coherent light beam through the ensembles and measuring the spin difference between them \cite{julsgaard2001experimental,aristizabal2021quantum}.  
In most experiments to date, the entanglement is produced in a Holstein-Primakoff regime, where only small amounts of entanglement are generated.  Schemes to make maximally entangled spinor state (\ref{eprz})
has been proposed using a sequence of QND measurements  \cite{chaudhary2023macroscopic}. For a more general spinor state, imaginary time evolution methods \cite{mao2022measurement} can be used, where the ground state of a given Hamiltonian can be found using a measurement-feedback scheme.  
For example, to make a general bipartite spinor state, the Hamiltonian (\ref{twospinorham}) is used in a imaginary time approach using QND measurements such as in Ref. \cite{kondappan2023imaginary}.  Then local unitary pulses $ V_1 $ and $ V_2 $ are applied to each ensemble individually.  This can be done in atom chips by optical Raman pulses. 

The scalability issues that face spin coherent states are improved for spinor states. The fundamental reason is that spin coherent states are symmetric under interchange of the whole molecule, while spinor states are only symmetric under local particle interchange. This means that operators that are symmetric under local interchange can generate spinor states, which is the case for both the QND interactions and the local unitary rotations.  Thus increasing the number of sysbsystems $ M $ can then by achieved by simply adding further atomic ensembles. This is analogous to how scalability is achieved in a modern qubit-based quantum computers, where additional qubits are augmented and entangled together.  This is more difficult for spin coherent states (Fig. \ref{fig1}(b)), where locally symmetric operators cannot generate the spin coherent states, and therefore one must spectroscopically control the entire molecular system. The exponential number of states which must be all controlled together makes it more difficult from a scalability point of view.

\section{Summary and conclusions}

We have analyzed the multipartite extension of spin coherent states.  Two natural generalizations were performed, either by a tensor product duplication (\ref{multiscs}), or by generalizing  the wavefunction in the bosonic form (\ref{multiscsbosonic}).  The former, which we call the multipartite spin coherent state, is symmetric under full interchange of the molecules, i.e. the states that are duplicated.  The latter, which we call the multipartite spinor state, is symmetric under local particle interchange.  These different symmetry properties lead to different properties of the state, as illustrated by the different dependence of expectation values and covariances of the bipartite spinor state.  Despite this, the correlations (in the sense of (\ref{corrdef})) of the maximally entangled bipartite spinor states were found to be identical to that of spin coherent states, taking an EPR form. Under single particle decoherence, symmetric observables have a dependence on decoherence that is at the same level as their corresponding microscopic versions, for both classes of states.  Their duplication can be taken advantage of with a higher signal-to-noise for symmetric observables, and is compatible with digital error correction strategies to reduce logical errors exponentially for classical readouts.  

Due to the locally addressable nature of spinor states, and their favorable properties with errors, these offer an interesting way of storing quantum information.  A scheme for quantum computation based on spinor states was discussed previously in Refs. \cite{byrnes2012macroscopic,byrnes2015macroscopic,byrnes2021quantum,abdelrahman2014coherent}.  One of the challenges of the scheme has been to find ways of mapping existing quantum algorithms intended for qubits to the spinor quantum computing framework in a experimentally feasible way.  The challenge here is to perform the quantum computation in the same way as with qubits but in spinor form, using experimentally reasonable manipulations.  We do not discuss this aspect in this paper, but we note that this has been achieved for several algorithms to date \cite{byrnes2015macroscopic,semenenko2016implementing,mohseni2021error}. Beyond the passive error correction approach that was discussed in Sec. \ref{sec:error}, there is some evidence that some natural error self-correction occurs naturally in atomic systems through the identical self rephasing effect (ISRE) \cite{deutsch2010spin}. A more detailed discussion of the exact way of a general quantum algorithm can be implemented is left as future work.

\section{Acknowledgments}
This work is supported by the National Natural Science Foundation of China (62071301); NYU-ECNU Institute of Physics at NYU Shanghai; Shanghai Frontiers Science Center of Artificial Intelligence and Deep Learning; the Joint Physics Research Institute Challenge Grant; the Science and Technology Commission of Shanghai Municipality (19XD1423000,22ZR1444600); the NYU Shanghai Boost Fund; the China Foreign Experts Program (G2021013002L); the NYU Shanghai Major-Grants Seed Fund; Tamkeen under the NYU Abu Dhabi Research Institute grant CG008; and the SMEC Scientific Research Innovation Project (2023ZKZD55).

\appendix
\section{Equivalence between bosonic and distinguishable states}
\label{app:equiv}

Consider the multinomial expansion of the unipartite spinor state (\ref{scsdefboson}), which is written as
\begin{align}
|\psi \rangle \rangle = &  \frac{1}{\sqrt{N!}} \sum_{k_0 = 0}^N \dots \sum_{k_{L-1} = 0}^N \binom{N}{k_0, \dots, k_{L-1}} \psi_0^{k_0} \dots  \psi_{L-1}^{k_{L-1}} \nonumber \\
& \times (a_0^\dagger)^{k_0} \dots (a_{L-1}^\dagger)^{k_{L-1}} | \text{vac} \rangle  .
\end{align}
Then using the definition (\ref{multifockone}) for the normalized Fock state we obtain (\ref{unispinorexpan}). 

Now let us compare this to the expanded spin coherent state (\ref{expandedscs}).  Matching the states with the same coefficient $ \psi_1^{k_1} \dots \psi_L^{k_L}$ gives the equivalence between the bosonic and distinguishable state versions
\begin{align}
& | k_0, \dots, k_{L-1} \rangle \leftrightarrow 
\frac{1}{\sqrt{\binom{N}{k_0, \dots, k_{L-1}}}} \nonumber \\
& \times \sum_{ q } P_q
| \underbrace{0 \dots 0}_{k_0} \underbrace{1 \dots 1}_{k_1} \dots \underbrace{L-1 \dots L-1}_{k_{L-1}} \rangle ,
\end{align}
where the state within the sum contains $k_l$ of the states in the state $ | l \rangle $, for $ l \in [0,L-1] $.  $ P_q$ is a permutation operator that interchanges any two of the $ N $ particles, and the sum runs over all distinct permutations.  There are a multinomial $ \binom{N}{k_0, \dots, k_{L-1}} $ number of such distinct permutations, giving the normalization factor. We note that the $L = 2$ version of this was discussed in Ref. \cite{byrnes2021quantum}.  

Counting the number of possible Fock states (\ref{multifockone}) with a total particle number $ N $ gives
\begin{align}
\sum_{k_0=0}^N \sum_{k_1=0}^{N-k_0} \sum_{k_2=0}^{N-k_0-k_1} \dots \sum_{k_{L-2}=0}^{N-\sum_{l=0}^{L-3} k_l } 1 = \binom{N+L-1}{L-1} . 
\end{align}

\section{Expectation values and covariances of the two qubit spinor state}
\label{app:expectation}

Here we give details of the evaluation of expectation values for the state  (\ref{schmidttwoqubit}).  We work in the Schmidt basis 
\begin{align}
| \tilde{\Psi} \rangle \rangle & = \frac{1}{\sqrt{ {\cal N}_\Psi }} \left(  \cos \chi a^\dagger_1 a^\dagger_2 + \sin \chi b^\dagger_1 b^\dagger_2 \right)^N | \text{vac} \rangle \nonumber \\
& = \frac{N! }{\sqrt{ {\cal N}_\Psi }} \sum_{k=0}^N \Psi_k |k \rangle_1 | k \rangle_2 ,
\label{fockexp}
\end{align}
where we defined
\begin{align}
\Psi_k = \cos^k \chi \sin^{N-k} \chi
\end{align}
and 
\begin{align}
| k \rangle_m  = \frac{ (a_m^\dagger)^{k}  (b_m^\dagger)^{N-k}  }{\sqrt{k! (N-k)! }} | \text{vac} \rangle  .  
\end{align}

\subsection{Single spin expectation values}

All expectation values are evaluated all in the Fock basis.  For example, the expectation value of $ \tilde{S}^z_m $ for $ m \in \{ 1,2 \}$ is
\begin{align}
&  \langle \tilde{S}^z_m \rangle = \langle \langle  \tilde{\Psi} | \tilde{S}^z_m |   \tilde{\Psi}  \rangle \rangle \nonumber \\
& = \frac{(N!)^2}{ {\cal N}_\Psi } \sum_{k=0}^N | \Psi_k |^2 (2k - N) \nonumber \\
& = \frac{1 }{\cos 2 \chi (\cos^{2N+2} \chi - \sin^{2N+2} \chi) }  \nonumber \\
& \times \Big[ N(\cos^{2N+4} \chi - \sin^{2N+4} \chi) \nonumber\\
&  + (N+2) \sin^2 \chi \cos^2 \chi (\sin^{2N} \chi -\cos^{2N} \chi) \Big]. 
\label{exactsz}
\end{align}
For large $ N $, we may drop terms that are not proportional to $ N $ and we obtain
\begin{align}
\langle \tilde{S}^z_m \rangle \approx \frac{N (\cos^{2N+2} \chi + \sin^{2N+2} \chi) }{\cos^{2N+2} \chi - \sin^{2N+2} \chi } \approx N \text{sgn}(\cos 2 \chi) ,
\label{signapprox}
\end{align}
where we made a further approximation that if  $ |\cos \chi| > |\sin \chi| $, the high powers will make the sine term negligible, and vice versa.  

The expectation value of $ \tilde{S}^x_m,  \tilde{S}^y_m $ are zero since these shift the Fock number of one of the subsystems by one in (\ref{fockexp}).

\subsection{Two-spin expectation values}

The expectation values of $ \tilde{S}^z_1 \tilde{S}^z_1 $ and $ (\tilde{S}^z_m)^2 $ are non-zero due to the correlated nature of the Fock states in the $ z $-basis.  We evaluate this as
\begin{align}
 \langle \tilde{S}^z_1 \tilde{S}^z_2 \rangle  = & \langle (\tilde{S}^z_m)^2 \rangle \nonumber \\
 = & \frac{(N!)^2}{ {\cal N}_\Psi } \sum_{k=0}^N | \Psi_k |^2 (2k - N)^2 \nonumber \\
 = & \frac{4 +2N +N^2 + N(N+2) \cos 4 \chi}{2 \cos^2 2 \chi} \nonumber \\
& - \frac{2(N+1)(\cos^{2N + 2} \chi +\sin^{2N + 2} \chi ) }{\cos 2 \chi (\cos^{2N+2} \chi - \sin^{2N+2} \chi) }  .
\label{exactszsz}
\end{align}
We may again approximate this expression for large $ N $ by dropping all terms except for those proportional to $ N $ or $ N^2$.  We then have
\begin{align}
 \langle \tilde{S}^z_1 \tilde{S}^z_2 \rangle  =  \langle (\tilde{S}^z_m)^2 \rangle \approx N^2 + 2N \left( 1- \frac{1 }{ |\cos 2 \chi |} \right),
 \label{szszapprox}
\end{align}
where we made a similar approximation as (\ref{signapprox}).

The expectation values of off-diagonal two-spin expectation values are zero except for 
\begin{align}
&  \langle \tilde{S}^x_1 \tilde{S}^x_2 \rangle  = -  \langle \tilde{S}^y_1 \tilde{S}^y_2 \rangle = 2  \langle \tilde{S}^+_1 \tilde{S}^+_2 \rangle \nonumber \\
 & =  \frac{(N!)^2}{ {\cal N}_\Psi }  \sum_{k=0}^N k (N-k+1) \Psi_k \Psi_{k-1} \nonumber \\
 & = \frac{1 }{2 \cos^2 2 \chi (\cos^{2N+2} \chi - \sin^{2N+2} \chi) }   \nonumber \\
&  \times \Big[ N \sin 4 \chi  (\cos^{2N+2} \chi + \sin^{2N+2} \chi) \nonumber \\
& - \sin^3 2 \chi  (\cos^{2N} \chi - \sin^{2N+2} \chi)  \Big] ,
\label{sxsxform}
\end{align}
where $ \tilde{S}^{\pm}_m = (\tilde{S}^x_m \pm i \tilde{S}^y_m)/2 $ and its application on the Fock states is given by $ \tilde{S}^+ | k \rangle = \sqrt{(k+1)(N-k)} | k+1 \rangle $ and $ \tilde{S}^- | k \rangle = \sqrt{k(N-k+1)} | k-1 \rangle $ \cite{byrnes2021quantum}.  The approximation for large $ N $ is obtained by keeping only terms proportional to $ N $, giving
\begin{align}
\langle \tilde{S}^x_1 \tilde{S}^x_2 \rangle  = -  \langle \tilde{S}^y_1 \tilde{S}^y_2 \rangle \approx N \text{sgn}(\cos 2 \chi) \tan 2 \chi
\end{align}
where we made a similar approximation as (\ref{signapprox}).  

Finally, we have 
\begin{align}
& \langle (\tilde{S}^x_m)^2 \rangle = \langle (\tilde{S}^y_m)^2 \rangle = \langle \tilde{S}^+_m \tilde{S}^-_m \rangle + \langle \tilde{S}^-_m \tilde{S}^+_m \rangle \nonumber \\
& =  \frac{(N!)^2}{ {\cal N}_\Psi }  \sum_{k=0}^N \Big[ k (N-k+1) + (k+1) (N-k) \Big] \Psi_k^2 \nonumber \\
 & = \frac{1 }{\cos^2 2 \chi (\cos^{2N+2} \chi - \sin^{2N+2} \chi) }   \nonumber \\
&  \times  \Big[ N(\cos^{2N+4} \chi - \sin^{2N+4} \chi) \nonumber \\
& + (N+2) \sin^2 \chi \cos^2 \chi (\sin^{2N} \chi -\cos^{2N} \chi) \Big] .
\end{align}
Using a similar approximation to  (\ref{signapprox}), we obtain
\begin{align}
\langle (\tilde{S}^x_m)^2 \rangle = \langle (\tilde{S}^y_m)^2 \rangle
\approx \frac{N}{| \cos 2 \chi |} . 
\end{align}

\subsection{Covariances}

The covariance matrix elements are evaluated using the definition (\ref{covmatrix}) and the above expressions.  The exception is the approximate expressions for $ \text{Cov} (\tilde{S}^z_1, \tilde{S}^z_2 ) $ and $ \text{Var} (\tilde{S}^z_m) $, where combining (\ref{signapprox}) and (\ref{szszapprox}) does not lead to a good approximation.  Instead, we use the exact expressions (\ref{exactsz}) and (\ref{exactszsz}) and collect the expressions in powers of $N$.  Terms propotional to $ N $ and $ N^2 $ are found to be small for large $N $, except at $ \chi = \pi/4 $.  Neglecting these terms and approximating the remaining terms gives the approximation
\begin{align}
\text{Cov} (\tilde{S}^z_1, \tilde{S}^z_2 ) = \text{Var} (\tilde{S}^z_m) \approx \tan^2 2 \chi .
\end{align}


\providecommand{\noopsort}[1]{}\providecommand{\singleletter}[1]{#1}%

\end{document}